# Intermediate phase, network demixing, boson and floppy modes, and compositional trends in glass transition temperatures of binary $As_xS_{1-x}$ system

*Ping Chen, Chad Holbrook, P. Boolchand*
Department of Electrical and Computer Engineering, University of Cincinnati, Cincinnati, Ohio 45221-0030

*D. G. Georgiev*
Department of Electrical Engineering and Computer Science, University of Toledo, Toledo, OH 43606

*K. A. Jackson*
Department of Physics, Central Michigan University
Mt Pleasant, MI

*M. Micoulaut*
Laboratoire de Physique Theorique de la Matiere Condensée, University Pierre et Marie Curie, Boite 121, 4, Place Jussieu, 75252 Paris, Cedex05, France.

Abstract

The structure of binary $As_xS_{1-x}$ glasses is elucidated using modulated- DSC, Raman scattering, IR reflectance and molar volume experiments over a wide range (8% < x < 41%) of compositions. We observe a reversibility window in the calorimetric experiments, which permits fixing the three elastic phases; *flexible* at x < 22.5%, *intermediate* in the 22.5% < x < 29.5% range, and *stressed-rigid* at x > 29.5%. Raman scattering supported by first principles cluster calculations reveal that the observed vibrational density of states has features of both pyramidal (PYR) ($As(S_{1/2})_3$) and quasi-tetrahedral (QT) ($S=As(S_{1/2})_3$) local structures. The QT unit concentrations show a global maximum in the intermediate phase (IP), while the concentration of PYR units becomes comparable to those of QT units in that phase, suggesting that both these local structures contribute to the width of the IP. The IP centroid in the sulfides is significantly shifted to lower As content x than in corresponding selenides, a feature identified with excess chalcogen partially segregating from the backbone in the sulfides, but forming part of the backbone in selenides. These ideas are corroborated by the proportionately larger free volumes of sulfides than selenides, and the absence of chemical bond strength scaling of $T_g$s between As-sulfides and As-selenides. Low-frequency Raman modes increase in scattering strength linearly as As content x of glasses decreases from x = 20% to 8%, with a slope that is close to the floppy mode fraction in flexible glasses predicted by rigidity theory. These results show that floppy modes contribute to the excess vibrations observed at low frequency. In the intermediate and stressed rigid elastic phases low-frequency Raman modes persist and are identified as boson modes. Some consequences of the present findings on the optoelectronic properties of these glasses are commented upon.

I. INTRODUCTION

Network glasses have traditionally been modeled as continuous random networks (CRNs). The description has worked well for several stoichiometric oxides but less so for chalcogenides. Some chalcogenides macroscopically phase separate displaying bimodal $T_g$s, as in the case of the Ag-Ge-Se ternary[1,2], materials that serve as active media in programmable non-volatile memories used for portable electronics[3]. In bulk glasses of the Ag-Ge-Se ternary, the nature of the two glass phases has also been established by calorimetry[4,5] and electric force microscopy[6]. Furthermore, even in chalcogenides that display a single $T_g$, demixing of networks can occur on nanoscale as either monomers or even large clusters decouple from the backbone, and enrich their functionality in remarkable ways. It is in this respect that the case of the binary $As_xS_{1-x}$ glass system has received attention [7-12] over the years starting from the early work of Ward[13] at Xerox, who recognized $S_8$ rings to decouple in S-rich glasses. It is generally believed that As serves the role of cross-linking chains of sulfur to acquire a local 3-fold coordination as in a pyramidal (PYR) unit, $As(S_{1/2})_3$, and at the stoichiometric composition, x = 40%, a CRN of S-bridging pyramidal units is realized[14]. Close analogies in molecular structure of the stoichiometric glass with the layered structure of c-$As_2S_3$ (orpiment) have been drawn. The molecular structure of $As_2S_3$ glass has also been compared to that of $As_2Se_3$ glass. If these stoichiometric glasses were to form CRNs, then their mean coordination number, $r = 2.40$, and following the ideas of Phillips[15] and Thorpe[16], the count of Lagrangian bonding constraints/atom, $n_c = 3$, for both these glasses[17]. Their glass structures can then be expected to be optimally constrained, and to possibly self-organize[18] to produce rigid but unstressed structures. Such a picture appears to largely describe the structure of $As_2Se_3$ glass[19], a composition that is almost in the reversibility window of binary $As_xSe_{1-x}$ glasses[19]. On the other hand, the stoichiometric composition, $As_2S_3$, is far from being in the reversibility window of binary $As_xS_{1-x}$ glasses, as will be shown in the present work. The underlying differences of structure between As-sulfide- and selenide glasses are not confined to the stoichiometric composition; in fact they extend to all other compositions as well. Do these peculiarities of structure of As-S glasses occur in other group V and group IV sulfides? We comment on the issue as well.

Glass transition temperatures ($T_g$s) provide a good measure[20] of network connectivity. For glass networks possessing the <u>same</u> connectivity, $T_g$s usually scale with strengths of chemical bonds. For example, in comparing binary ($Ge_xX_{1-x}$ ( X = S, or Se)), and ternary ($P_xGe_xX_{1-2x}$ (X = S, or Se)) glasses, one generally finds the ratio of $T_g$s of sulfides to selenides to scale as the ratio of single chemical bond strengths[21]. However, such is not the case for the $As_xX_{1-x}$ binaries (X = S or Se). For example, $T_g$ of $As_2S_3$[22] (205 (2)°C) is almost the same as that of $As_2Se_3$ [8, 19] (197(2) °C ). If these glasses were to possess the <u>same</u> network connectivity, one would have expected $T_g$ of $As_2S_3$ to be 13% higher ( 255°C ) than of $As_2Se_3$, since single As-S covalent bonds[23] (47.25 kcal/mole) are 13% stronger than single As-Se covalent bonds (41.73 kcal /mole ). The absence of bond strength scaling of $T_g$ is not restricted to the stoichiometric composition, x =

40%, but it occurs at other As concentrations x as well. These data suggest that there are intrinsic differences of glass structure between As- sulfides and –selenides.

Differences of glass structure between As-sulfides and As-selenides have a bearing on infrared optical[24, 25] applications of these materials. Photodarkening in amorphous films of $As_2S_3$ and $As_2Se_3$ has been examined[26] using near band gap light (500 nm for a- $As_2S_3$ film, and 650 nm for a-$As_2Se_3$ film). The results show that not only band gap reduction but also rate of photodarkening is higher in the sulfide than in the selenide films. These glasses also display large optical nonlinearities, low phonon energies and low nonlinear absorption[27]. Material properties such as nonlinear changes in refractive index , electronic polarizability, molecular orientation, electrostriction and thermal effects are of relevance in this context[28]. Amongst these, fast electronic nonlinearity is of particular interest; it is determined by electron distribution in various local structures, and can be expected to change with glass composition. Thus, introduction of Se in As-S glasses increases the non-resonant refractive nonlinearity up to 400 that of fused silica[29]. In ternary Ge-As-Se glasses, it has been shown that substitution of Ge by As promotes while replacement of Se by As suppresses optical non-linearity[30]. Furthermore, refractive nonlinearity is found to vary with composition in over connected ($r$ >2.4) glasses with the nonlinearity being highest for the lowest Ge concentration[31]. The microscopic origin of these non-linear effects remains an open question in the field. It is likely that a basic understanding of glass structure would help in addressing the underlying issues.

In the present work we focus on compositional dependence of glass structure; we have examined the thermal (modulated-DSC) and optical ( Raman and IR reflectance) behavior of bulk $As_xS_{1-x}$ glasses over a wide range of compositions, 8% < x < 41%. In Raman scattering compositional trends in vibrational modes at low-frequency (0-100 cm$^{-1}$), bond-bending (100-250 cm$^{-1}$) and bond-stretching (250-500 cm$^{-1}$) regimes are analyzed. Results of first principles cluster calculations[32] are used to analyze the Raman and IR vibrational density of states. The present data suggest that in addition to 3-fold coordinated As, quasi-4-fold coordinated As local structures (with As having 3 bridging and one non-bridging S neighbor) are also present in the S-rich glasses. Both these local structures are found in binary P-S [21, 33, 34] and P-Se glasses[35, 36]. Our m-DSC experiments reveal the existence of a reversibility window[37] in the 22% < x < 29% range, an observation that fixes the three elastic phases- *flexible, intermediate and stressed-rigid* in the present As-S binary. The window is found to be shifted to lower As content (x) in relation to the one in corresponding selenides. We trace the observation to partial demixing of excess S from networks of the present sulfides but the complete mixing of excess Se in the corresponding selenides. The strongly excited low-frequency Raman modes, also called boson modes, are analyzed in the present sulfide glasses. In the elastically flexible regime (x < 20%) these low frequency vibrations are shown to come from *floppy modes*, while those in the intermediate and elastically stressed-rigid glasses probably come from inter cluster soft modes.

In sections II we present some background considerations to understanding compositional trends in $T_g$. Results of first principle cluster calculations to analyze the Raman vibrational modes are presented in section III. Experimental results and their discussion in relation to issues of glass molecular structure, floppy modes, boson modes, and the three elastic phases in the present As-S binary appear in sections IV and V respectively. A summary of the present findings appears in the conclusions.

## II. COMPOSITIONAL TRENDS IN $T_g$ AND GLASS STRUCTURE

**A. Stochastic agglomeration and network connectivity.**

A useful means to understanding $T_g$ variation with glass chemical composition or network structure is provided by a stochastic agglomeration theory (SAT). The theory allows a prediction of $T_g$ with modifier composition when agglomeration of atoms forming part of a glass-forming liquid proceeds in a *stochastic fashion*. The central idea of the theory[20, 38, 39] is to relate an increase of viscosity to network formation as chemical bonds between well-defined *local structures* are formed at random upon cooling. For a base glass composed of A-atoms, which is modified by alloying B-atoms, i.e., an $A_xB_{1-x}$ binary system, theory predicts[20, 38, 39] a parameter-free slope of $T_g$ with glass composition x in the limit when x is low,

$$\left[\frac{dT_g}{dx}\right]_{x=0, T_g=T_{g0}} = \frac{T_{g0}}{\ln\left[\frac{r_B}{r_A}\right]} \quad (1)$$

In equation (1), $r_B$ and $r_A$ are the coordination numbers of the additive atom (B) and the atom (A) comprising the base glass network, and $T_{g0}$ represents the glass transition temperature of the base material. The denominator appearing in equation (1) gives an entropic measure of network connectivity[40], and the lower the entropy the higher the slope of $T_g$ with x. For the case $SiO_2$, which has a large $T_{g0} \sim 1200°C$, a large slope $dT_g/dx$ is expected and indeed found in sodium silicate glasses in the low modified regime. For the case of $As_xSe_{1-x}$, ($r_A=2$) and if we take As atoms to be 3-fold coordinated ($r_B=3$), one expects the slope to be 6.17°C/mol.% of As. The observed slope is found[36] to be 4.1°C/mol.% of As. We have suggested in earlier work[5] that presence of 4-fold coordinated As as in a QT unit will increase the number of ways to connect arsenic and selenium atoms together, and will thus decrease the slope $dT_g/dx$, and permit one to reconcile the observed slope dTg/dx if the concentration of such units is in the 20 to 30% range.

Elemental Se is a good glass former composed largely of polymeric chains of selenium. On the other hand, elemental S melts upon cooling condense into a molecular solid composed of $S_8$ rings. These considerations help us reconcile why binary (Ge or As)$_x$Se$_{1-x}$ melts give rise to homogeneous glasses extending all the way to pure Se, while corresponding sulfide melts, in general, segregate into a glassy backbone and S$_8$ crowns as x decreases to ~0.20. And as x approaches 0, melts spontaneously crystallize. The present As$_x$S$_{1-x}$ binary glasses intrinsically segregate on a molecular scale as x approaches 0, so that SAT as we know it cannot be directly applied. Inferring aspects of glass structure at low x using SAT poses new challenges. Fortunately, signature of S$_8$ rings segregating from the backbone in sulfide glasses come independently from Raman scattering and modulated DSC experiments as will be illustrated in the present work. Many of these ideas were known more than 25 years ago. What is new, however, is that use of m-DSC has now permitted establishing scan rate independent glass transition temperatures ($T_g$s). At higher x when network formation is well developed, trends in $T_g$ can be understood in terms of SAT. These thermal data complemented by optical ones (Raman and IR reflectance measurements), afford new insights into the molecular structure of the present sulfide glasses as we shall illustrate in this work.

**B. Chemical bond strength rescaling of glass Transition temperatures**

The importance of network connectivity in determining the $T_g$ of network glasses has evolved elegantly from SAT. The importance of chemical bond strengths in determining $T_g$ was emphasized by Tichý and Tichá[41] from empirical considerations. It is becoming transparent that bond-strength scaling of $T_g$s does occur but only if the underlying network structures being compared possess the same connectivity. An illustrative example is the case of the Ge$_x$P$_x$X$_{1-2x}$ ternaries with X = Se or S (Figure 1). A perusal of the data[42, 43] shows that in the 10% < x < 18% range, $T_g$ - ratio of sulfide to Selenide glasses is found to be 1.13, and the ratio is found to be *independent* of x. The finding illustrates that connectivity of underlying backbones steadily increases as the concentration x of the cross-linking atoms, P and Ge increases in both these systems. We understand the scaling factor of 1.13 in terms of the higher chemical bond strength of Ge-S, and P-S bonds in relation to Ge-Se and P-Se ones. The Pauling single bond strength data[23] on the underlying bonds is as follows: Ge-S bond: $E_b$= 55.52 kcal/mole , Ge-Se bond: $E_b$ = 49.08 kcal/mole, P-S bond: $E_b$ = 54.78 kcal/mole, P-Se bond : $E_b$ = 49.72 kcal/mole. These data yield the ratio of chemical bond strengths of the P and Ge sulfide bonds to corresponding selenide bonds of 1.12. These ideas on bond-strength rescaling of $T_g$s will be useful when we visit results on the present As-S glasses and compare them to those on corresponding selenides.

At x < 10%, a re-scaling of $T_g$s no longer holds since in the sulfide glasses $S_8$ rings steadily decouple leading to loss of a network, while in the selenides a network structure continues to persist even as x approaches 0. The signature of $S_8$ ring decoupling is the appearance of the $T_\lambda$ transition (figure 1). $T_g$s in the 10% < x < 0% range decrease almost linearly to extrapolate to a value of about -50°C as x approaches 0. We assign the extrapolated $T_g$ = -50°C to that of a S-ring glass. On the other hand, if one linearly extrapolates the $T_g(x)$ data from x ~ 10% backwards to x = 0, one obtains an extrapolated $T_g$ of about 100°C. We assign the extrapolated $T_g$ of 100°C to that a $S_n$-chain glass. These extrapolated $T_g$ values of a S-chain and a S-ring glass are independently corroborated by results on several other binary and ternary sulfide glass systems.

### III.  FIRST PRINCIPLES CLUSTER CALCULATIONS OF VIBRATIONAL DENSITY OF STATES.

To investigate the origin of key features in the Raman spectra of $As_xS_{1-x}$ glasses, we conducted first-principles calculations based on density functional theory in the local density approximation[44] (LDA). The calculations were performed using the NRLMOL code.[45, 46] Extensive Gaussian basis sets were used to represent the electronic orbitals. Pseudopotentials[47] were used for the As and S atoms, while H atoms were treated[32] in an all-electron format.

Attention was focused on two structural building blocks of the glasses, the As-$S_3$ pyramid (PYR) and the quasi-tetrahedral (QT) As-$S_4$ unit, both shown in Fig. 2. Hydrogen atoms were added to the three basal S atoms in each of these models to tie off dangling bonds so as to better mimic the local chemistry of these units in the glass. The S-H bonds nominally represent S-As bonds to the glass network. The Pauling electronegativities of As and H are 2.18 and 2.20, respectively, so that one expects S-H and and S-As bonds have very similar polarities. The cluster bond lengths and bond angles were relaxed to a minimum energy configuration and the vibrational normal modes of the clusters were calculated using a standard technique.[48] IR intensities and Raman activities for the vibrational modes were then calculated within the LDA. The method has been described fully elsewhere.[48] Briefly, IR and intensities and Raman activities can be related to changes in cluster dipole moments and polarizabilities caused by atomic displacements in the vibrational modes. These changes can be computed directly within the LDA, using no adjustable parameters or empirical input. Calculated results for a diverse set molecules have been shown to be in good agreement with experimental measurements.[48]

The results for the PYR and QT units are given in Table I. Mode frequencies and total IR intensities and Raman activities for key vibrational modes are listed. Normal mode eigenvectors indicating the pattern of atomic displacements in each mode are shown in Fig. 2. It is interesting to note that the symmetric and asymmetric modes associated with the PYR unit are nearly degenerate, at 352 and 355 cm$^{-1}$, respectively, while the analogous modes in the QT unit are split, at 335 and 365 cm$^{-1}$. (The asymmetric stretch modes in both models are doubly degenerate.

Only one set of eigenvectors is shown for each model in Fig. 2. The QT unit also features a mode at 537 cm$^{-1}$ that arises from a stretch of the bond between As and the one-fold coordinated S atom. This mode clearly has no analogue in the PYR model and is well-separated from the remaining As-S stretch modes. The results, presented in Table I and Fig. 2, are discussed further below.

## IV. EXPERIMENTAL

### A. Sample synthesis

Figure 3 shows the phase diagram of the As-S binary system[49, 50]. In the phase diagram, the crystalline phases shown are crystalline $S_8$ that exist in three forms[51] (α, β and γ), c-$As_2S_3$ (orpiment)[52], c-$As_4S_4$ (realgar) that exists in two forms[53] (α, β) and c-$As_4S_3$ (dimorphite) that exists in three forms[50] (α, β and γ). Both crystalline $As_4S_4$ and $As_4S_3$ are molecular crystals composed of monomers. A perusal of the $As_xS_{1-x}$ phase diagram shows that there is a eutectic near x ~ 1%. The glass forming range in the $As_xS_{1-x}$ binary extends from about 5% < x < 55% range. At x < 5% melts are largely composed of $S_8$ rings, while at x > 55% melts crystallize into Realgar or $As_4S_3$ monomers. Other crystalline forms found in mineral data include $As_4S$ (duranusite)[54] and $As_4S_5$ (uzonite)[55].

Glass samples of typically 2 grams in size were synthesized using 99.999% $As_2S_3$ and elemental S pieces from Cerac Inc. The starting materials were weighed and mixed in a $N_2$ gas purged glove bag, and then sealed in evacuated (< 10$^{-7}$ Torr) *dry* quartz tubings of 5.0 mm id and 1mm wall thickness. Temperature of quartz tubes was slowly increased to 650°C at a rate of 1°C/min. Melts were homogenized by mixing them at 600°C for 2 days or longer (see below). Glass samples were realized by quenching melts from 50°C above the liquidus[49] into cold water.

Once quenched, glass sample homogeneity was ascertained by FT-Raman scattering along the length of sealed quartz tubes used to synthesize samples. Periodic evaluation of samples with increasing reaction time of melts at 650°C revealed that Raman lineshapes became indistinguishable only after melts were homogenized for typically 2 to 3 days. Once samples were synthesized, these were stored in evacuated Pyrex ampoules to avoid hydrolysis.

### B. Molar volumes

An 8 inch long quartz fiber suspended from a pan of digital balance (Mettler model B154) and a hook to support a sample at the far end was used to measure mass densities using the Archimedes principle. Glass samples of typically 100 mgs or larger in size were weighed in air and in pure ethyl alcohol. The density of alcohol was calibrated using a single crystal bulk piece of Si and Ge as density standards. With the arrangement, we could obtain the mass density of samples to an accuracy of 3/4% or less. Resulting molar volumes as a function of glass composition are summarized in Figure.4 and show $V_m(x)$ decreases with increasing x but with a local minimum in the 20% < x < 28% range, the reversibility window (see below). For comparison we have also included in Figure 4, molar volumes on corresponding $As_xSe_{1-x}$ glasses taken from the work of Georgiev et al.[19] We shall return to discuss these results later.

### C. Modulated – Differential Scanning Calorimetry

Glass transition properties were examined using a modulated DSC, model 2920 from T.A. Instruments. Typically about 30 mgs quantity of a sample, hermetically sealed in Al pans, was heated at a scan rate of 3°C and a modulation rate of 1°C/100sec. A scan up in temperature across $T_g$ was then followed by a scan down in temperature. Scans of pure S and those of S-rich binary glasses at x = 8 and 15% are reproduced in Figure 5. In pure S (Figure 5a), we observe 3 endothermic events; (i) an α-β solid solid phase transformation near 108.6°C (ii) a melting of β –

S near 118°C and followed by (iii) a $T_\lambda$ transition near 160.6°C. The $T_\lambda$ transition is identified with opening of $S_8$ rings to form polymeric S-chains leading to the formation of viscous or plastic S [56]. In the inset of Figure 5a, the scan of pure S is enlarged, and it highlights the non-reversing enthalpy associated with the $T_\lambda$ transition, and one observes an endotherm exclusively. In S-rich glasses, and particularly at low x (< 23%), one observes in general, the *total heat flow* scan to reveal two thermal events; a glass transition endotherm in the 20° C < T < 125°C range, followed by a complex heat flow profile associated with the $T_\lambda$ transition. Help in understanding the complicated total heat flow profile comes from examining the deconvoluted *reversing* and *non-reversing* heat- flow profiles. In the *reversing heat flow* scan, one observes a glass transition endotherm to display a rounded step in $C_p$, and one fixes the $T_g$ value by the inflexion point of the step. The $T_\lambda$ transition also displays a rounded step in $C_p$, and one fixes the transition temperature by the inflexion point of the step. The *non-reversing heat* flow, as expected, shows a Gaussian like endotherm for the $T_g$ event, but for the $T_\lambda$ transition one observes an exotherm followed by an endotherm as illustrated in Figures 5b and 5c. These data suggest that as T increases, $S_8$ rings first become mobile near about 100°C and coalesce to form nano-crystalline fragments, a process that releases heat (exotherm). With a further increase of T, these fragments next open and are incorporated in the backbone, events that contribute to the endotherm. And as the As content of the glasses " x " exceeds 15%, $T_g$s steadily increase while the strength of the $T_\lambda$ transitions steadily diminishes as both $\Delta C_p$ and the non-reversing heat flow term $\Delta H_{nr}$ decreases ( Figure 6a and 6b). These data suggest that the process of alloying As in S leads to growth of network backbone as concentration of $S_8$ rings steadily decreases, and at x > 25%, there is little or no evidence of $S_8$ rings present in glasses. Even in the most S-rich glass investigated here, we find no evidence of the α→ β transition as seen in pure S, suggesting that $S_8$ rings formed in our glass samples do <u>not</u> precipitate to form microcrystalline clusters of α – sulfur. These $S_8$ crowns, most likely, are randomly distributed with the network backbone. Trends in $T_g$ for present $As_xS_{1-x}$ glasses appear in Figure 7, where we show for comparison $T_g$ s in binary $As_xSe_{1-x}$ glasses. We shall discuss these results later.

In an m-DSC experiment, deconvolution of the heat flow into a reversing and a non-reversing heat flow has profound consequences in understanding the nature of glass transition[37]. The former heat flow relates to ergodic events such as changes in vibrational motion, while the latter one relates to non-ergodic events such as changes in configurations accompanying structural arrest near $T_g$ as a glass softens. Data on chalcogenide glasses broadly reveals three generic types[57] of glass transitions based on connectivity of their backbones. Weakly cross-linked backbones form elastically *flexible* networks, which display a non-reversing enthalpy term, $\Delta H_{nr}$ (T), that is Gaussian-like, symmetric and narrow ($\Delta T \sim 20°C$) in width and the term ages. Highly crosslinked backbones form elastically *stressed-rigid* networks, which display a $\Delta H_{nr}$ (T) term that is broad ($\Delta T \sim 40°C$) and asymmetric with a high T tail and the term ages. On the other hand, optimally crosslinked backbones form elastically rigid but unstressed networks (Intermediate Phase) which display a $\Delta H_{nr}$ (T) term that is not only minuscule but also does not age much. The vanishing of the $\Delta H_{nr}$ term for IP glass compositions suggests that such networks possess liquidlike entropies, a feature that derives[58] from the presence of at least two isostatic local structures.

It is usual to make a frequency correction to the $\Delta H_{nr}$ term by subtracting the $\Delta H_{nr}^{cool}$ term measured in a cooling cycle from the $\Delta H_{nr}^{heat}$ term measured in the heating cycle.

$$\Delta H_{nr}^{freq.\ corr.} = \Delta H_{nr}^{heat} - \Delta H_{nr}^{cool} \qquad (2)$$

The frequency corrected $\Delta H_{nr}^{freq.\ corr.}$ term is independent of the scan rate employed. Thus, for example, the $\Delta H_{nr}^{heat}$ and $\Delta H_{nr}^{cool}$ terms for a sample at x = 28% are 0.86 cal/g and 0.73 cal/g respectively, upon a frequency correction $\Delta H_{nr}^{freq.\ corr.}$ is found to be 0.13 cal/g (Fig.8). The observed trends in non-reversing enthalpy reveal a narrow, deep but sharp reversibility window that onsets near x = 22.5%, and ends near x = 29%, (Figure 8). For convenience, the reversibility window in As-Se glasses [22] is also included in Fig.8. We find the window in sulfide glasses is shifted to lower x in relation to the window in selenide glasses.

### D. Raman scattering

Raman scattering was studied using a dispersive T64000 triple monochromator system from Horiba Jobin Yvon Inc, using 647 nm excitation. The dispersive system made use of a microscope attachment with a 80X objective, bringing laser light to a fine focus (2 μm spot size) and the scattered radiation detected using a CCD detector. Samples in a platelet form were encapsulated in a MMR Joule-Thomson refrigerator to avoid photo-oxidation and heating. Spectra could be recorded from 3 cm$^{-1}$ to 600 cm$^{-1}$ range permitting the boson peak to be investigated. Typically 2 mW of laser power was brought on the microscope table over a 2 μm spot size. Separately samples were also studied using a Thermo-Nicolet FT Raman system with 1.06μm excitation from a Nd-YAG laser. The observed lineshapes in either set looked very similar. In the FT-Raman experiments samples were encapsulated in evacuated quartz tubing and the laser beam typically 260 mW power was brought to a loose focus (150 μm spot size) onto the samples. Figure 9 gives a summary of the observed lineshape using an FT Raman system for glass compositions in the 8% < x < 41% range. These results are quite similar to those reported earlier[8, 11, 59, 60] by several groups. At low x (< 25%) one observes sharp modes that are readily identified with sulfur rings and chains. In the mid-x range (25% < x < 40%), the lineshape is dominated by a broad band in the $300 < \upsilon < 400$ cm$^{-1}$ range. At high x (>40%), one begins to observe again some sharp features labeled as R in the spectra (Figure 9). These R-modes were identified earlier by Georgiev et al.[22] as belonging to those of Realgar ($As_4S_4$) molecules that demix from network structure once x approaches x> 38%.

Guided by first principles cluster calculations (section III), we are now able to better to decode the vibrational density of states and understand the structure consequences. The mid-x region is of interest, particularly since it is in this range one observes the RW. Examples of Raman lineshape deconvolution for samples at x = 8%, 10% and 15% appear in Figure 10.

We concur with the earlier assignments[61] of the narrow modes near 151 cm$^{-1}$ ($S_8$), 217 cm$^{-1}$ ($S_8$), 440 cm$^{-1}$ (second order scattering from 217 cm$^{-1}$), 461 cm$^{-1}$ ($S_n$ chains) and 474 cm$^{-1}$ ($S_8$) as belonging to modes of either $S_8$ rings[61] or $S_n$ chains. The cluster calculations place the symmetric ($335^{cal}$ cm$^{-1}$) and asymmetric stretch ($365^{cal}$ cm$^{-1}$) of quasi tetrahedral S=As($S_{1/2}$)$_3$ units (Table 1). We thus assign the modes observed near 333 cm$^{-1}$ and 380 cm$^{-1}$ respectively to these vibrational modes of QT units. The As=S stretch mode of such a unit, we believe contributes to the mode near 500 cm$^{-1}$ in the spectra (Table1). The cluster calculations place the symmetric ($352^{cal}$ cm$^{-1}$) and asymmetric ($355^{cal}$ cm$^{-1}$) stretch of pyramidal As($S_{1/2}$)$_3$ units quite close to each other, and we assign the feature observed near 365 cm$^{-1}$ to both these vibrational modes.

The observed FT-Raman lineshapes in the 250 cm$^{-1}$ to 550 cm$^{-1}$ range were fitted using a least squares routine to a superposition of above-mentioned peaks. Lorentzian or Gaussian-Lorentzian sum functions are used for the sharper $S_8$ ring modes at 474 cm$^{-1}$ in S-rich samples and Gaussian functions were used for all other peaks. Linear baseline was safely used as we exclude the region below 250 cm$^{-1}$ which involve complicated baseline due to boson peak and holographic notch filter cutoff. Examples of lineshape fits at x= 8%, 10% and 15% appear in fig. 10.

From these fits, we have also extracted the normalized scattering strength ($A_\upsilon/A$) of the modes at $\upsilon_1$ = 333 cm$^{-1}$ (QT$^{ss}$), $\upsilon_2$ =365 cm$^{-1}$ (PYR) and $\upsilon_3$ =495 cm$^{-1}$ (QT), and figure 11 gives a plot of the compositional trends of mode scattering strengths. The total integrated area A under the various modes was used to normalize the individual mode scattering strengths. As expected, modes assigned to S-chains (461 cm$^{-1}$) and rings (474 cm$^{-1}$) monotonically decrease in scattering strength to nearly vanish as x increases to 40%. On the other hand, modes assigned to QT units (333 cm$^{-1}$ and 495 cm$^{-1}$) show a global maximum in the 22% to 29% range, compositions belonging to the RW. The scattering strength of the mode near 365 cm$^{-1}$ assigned to PYR units is found to increase monotonically as x approaches 40% with a mild plateau in the RW range. We consider these compositional trends in scattering strengths of Raman modes to be far more diagnostic than the mode frequencies alone, and this is a point we shall return to discuss later. Figure 12 summarizes the glass composition dependence of the Raman active bond-stretching mode frequencies due to QT and PYR units. One finds that in the 10% < x < 30% range mode frequencies steadily red-shift by nearly 15 cm$^{-1}$. This is unusual, and we shall comment on the behavior in the next section.

In the observed lineshapes, scattering in the low frequency range increases remarkably as the As content of glasses x < 20% as illustrated in figure 13, which gives the Dispersive Raman spectra. In making these plots we have normalized the observed spectra to the same laser power, and find that compositional trends in scattering strength of modes in the bond-stretching regime ( $S_8$ ring bond-bending mode (at 217 cm$^{-1}$), and the stretching mode of PYR units (at 365 cm$^{-1}$ and stretching modes of QT units) monotonically change as expected. To better visualize these excitations, we present in Figure 13, the experimental ($I_{expt}$) Raman lineshapes along with reduced Raman vibrational density of states ($I_{red}$) at several compositions. These data demonstrate that low frequency vibrational modes <u>dominate</u> Raman scattering ( Figure 13) in the present sulfides at low x , and particularly at x < 20%, a feature that was also noted by earlier workers in the field [60]. The boson mode in $As_2S_3$ glass was also observed in IR transmission experiments [62].

Raman scattering can, in general, be written[63] as follows,

$$I_{expt} \sim C(\omega)g(\omega)[n_B + 1]/\omega \qquad (3)$$

where $n_B$ represents the Bose occupation number, $C(\omega)$ the photon-vibration coupling constant and $g(\omega)$ the vibrational density of states (VDOS). To obtain the excess VDOS ( e-VDOS) over Debye-like vibrations, we follow the following procedure[64]. From the observed ( $I_{expt}$) Raman scattering, we obtain the reduced Raman DOVS,

$$I_{red} = I_{expt}/(n_B + 1) \qquad (4)$$

and subtract an estimated Debye-like density of states to obtain eVDOS, as illustrated for two glass compositions at x = 10% and 40% in Figure 14. In estimating the Debye density of states we scale the slopes of these curves as $1/v^3$, where v is the speed of sound[65]. The resulting excess (e)-VDOS are illustrated as the cross-hatched region in Figure14. A plot of the peak frequency, $\upsilon_{edos}$ and integrated intensity, $I_{edos}$ of the e-VDOS appears in Fig15 (a) and (b) respectively. One finds that $\upsilon_{edos}$ (x) (Figure 15a) smoothly increases from a value of 42 cm$^{-1}$ at x = 8% to 48 cm$^{-1}$ at x = 41%, but with a local minimum in frequency near x = 25% in the reversibility window. On the other hand, $I_{edos}$ (x) remains largely unchanged in the 41% < x < 20% range but increases almost linearly in the 20% < x < 8% range. We shall return to discuss these results in section V.

### E. IR reflectance

A Thermo-Nicolet model FTIR model 870 with a smart collector accessory was used to examine specular reflectance data from polished platelets of $As_xS_{1-x}$ glass samples. Typical sample size used was 5mm in diameter and 2mm thick. Reflectivity was studied over the 50 cm$^{-1}$ to 600cm$^{-1}$ range using solid state substrate beam splitter and DTGS detector with polyethylene window. A typical measurement used 600 scans with a 12 minutes scan time to give 4 cm$^{-1}$ resolution. A polished stainless steel surface was used as a reference for background. A Kramers Kronig transformation of the reflectance signal was performed with GRAMS software to generate the absorbance, the Transverse Optic (TO) and Longitudinal Optic (LO) IR response.

A summary of the reflectance signal data recorded at several glass compositions appears in Figure 16(a). These data are quite similar to the ones reported earlier[11]. The ir TO ($\varepsilon_2$) response for glasses is summarized in Figure 16(b), and is found to be quite similar to those reported earlier[11]. We have summarized the TO as well as the LO response at select glass compositions (20%, 30% and 40%) in Figure 17. The data for the stoichiometric composition, x = 40%, may be compared to earlier reports [11, 12]. In these data one can see vibrational features near 312 cm$^{-1}$, near 340 cm$^{-1}$, identified earlier[12] with the symmetric and asymmetric stretch of PYR units. In addition there are other features present in these data, which make the IR response consistent with the present and earlier[22] Raman data. For example, network modes of PYR units[22] must also contribute to the ir response particularly in the 360 to 400 cm$^{-1}$ range. In this range, presence of Realgar units ($As_4S_4$) in the $As_2S_3$ glass, noted earlier in Raman scattering, will also contribute to (features near 300 cm$^{-1}$ and 370 cm$^{-1}$) in the IR response. Mori et al.[11] observed a small but clearly resolved feature near 490 cm$^{-1}$ in the IR absorption measurements of the stoichiometric

glass( x = 40%). The feature comes from the stretch of As=S double bonds of QT units. The broad TO response observed in the 280 cm$^{-1}$ to 500 cm$^{-1}$ range ( Figure 17) , must then also include features from the symmetric and asymmetric stretch of the QT units.

In a glass of composition x = 30%, Raman scattering reveals no evidence of Realgar units. In the IR response one thus expects contributions from both PYR and QT units as shown in the middle panel of Figure 17(b). At x = 20%, Raman scattering reveals significant contribution from $S_n$ chains and $S_8$ rings. One thus expects, in addition to modes of QT and PYR units, those of $S_n$ chains and rings to contribute to the IR response near 480 cm$^{-1}$. One can see evidence of some IR response in the 440 to 500 cm$^{-1}$ range in the data of Figure 17(c). Unlike Raman modes that are reasonably sharp, the IR response from these glasses reveal modes that are quite broad, and a unique deconvolution of the observed lineshape becomes more challenging. Finally, a perusal of the data of Figure 17 illustrates the LO response to be shifted to higher frequency in relation to the TO response. In polar semiconductors internal electric fields add to the external one and blue shift the LO response to higher frequencies in relation to the TO response, as expected.

## V. DISCUSSION

### A. **Identification of the three elastic phases in binary $As_xS_{1-x}$ glasses**.

Reversibility windows (RWs) have been observed in chalcogenides [18, 19, 43, 66, 67] and more recently in oxide glasses[68] as well. These windows are identified with glass compositions belonging to an Intermediate Phase (IP), bordered on the low connectivity end by a *flexible* phase and on the high connectivity end by a *stressed-rigid* phase. Glass compositions in these windows form networks that are rigid but unstressed. Materials in the IP are functionally quite different from flexible or stressed-rigid phases; they age minimally and display dynamic reversibility, and have often been compared to proteins in the transition state that fold reversibly. These features of the IP are reviewed as representing a self-organized[69-72] phase of disordered matter.

In figure 15, we have assembled the compositional variation of $T_g(x)$, non-reversing enthalpy $\Delta H_{nr}(x)$, and molar volumes, $V_m(x)$ in the present binary glasses. The minuscule $\Delta H_{nr}(x)$, in the 22.5% < x < 29.5% range, serves to define the reversibility window (RW) in the present glasses. In analogy to earlier work on chalcogenides, we identify the RW with the IP of the present glasses. The variation of molar volumes reveals an almost linear decrease in the 10% < x < 40% range, but with some evidence of a <u>local</u> <u>minimum</u> in the RW (figure 18), surely the signature of space filling of the IP. Identification of the IP, thus permits fixing the three elastic regimes in the present binary glasses, compositions at x < 22.5% belong to the elastically *flexible* phase, while those on the high connectivity end at x > 29.5%, to the elastically *stressed rigid* phase. The stoichiometric glass composition, $As_2S_3$ (x = 40%) is not found to be part of the IP.

In the *flexible phase* of the present sulfides networks rapidly segregate as the As concentration decreases from 20% to 8% range. The almost linear reduction of $T_g(x)$ at x < 22% (Figure 15a), the increase in concentration of $S_8$ rings as reflected in growth in scattering strength of 217 cm$^{-1}$ (bending mode of $S_8$ rings) in Raman scattering (Figure 15a), and the appearance of the $T_\lambda$ transition in calorimetry, each data corroborates the general picture of $S_8$ crowns demixing from the backbone of glasses in the *flexible* phase. And even at the lowest As concentration examined here, we could not detect $S_8$ crowns segregating to form the crystalline orthorhombic S phase; there is no evidence of the α→β transition in calorimetry (Figure 5). In the IP, and particularly in 22.5% < x < 25% range, there is evidence of a small (2%) but decreasing concentration of $S_8$ rings present as x increases to about 25%.

In the *stressed rigid* phase, one finds $T_g(x)$ to increase with x, which is consistent with PYR

and QT units serving to crosslink $S_n$ chain segments and increasing connectivity of networks. The concentration of both PYR units increases with x in the 29% < x < 40% range. In spite of this trend, Raman vibrational mode frequencies of the two building blocks are found not to increase. These data suggest that the PYR and QT units and $S_n$ chains must <u>not</u> form part of the same fully polymerized structure as has been traditionally assumed. In several binary[66, 73] and ternary[67, 74] selenides that usually form fully polymerized networks, Raman scattering derived optical elastic constants (or mode frequency squared) display an increase as a power-law with glass composition x in the *stressed-rigid* phase. The absence of such a power-law variation underscores that underlying networks of the present sulfides are not fully connected (see below).

## B. Local structures of binary $As_xS_{1-x}$ glasses and the Intermediate Phase

Identification of the three elastic phases in the present binary begs the broader issue- what aspects of glass molecular structure control their elastic behavior? In particular, can one understand the width and centroid of the IP in the present binary sulfides? Several numerical approaches to model IPs [69-72] in the selenides have been discussed recently. In the sulfides that intrinsically segregate on nanoscale, experiments can however provide crucial insights on these phase separation effects. And such effects are likely to serve as important bounds in modeling IPs as well.

The IP in the present sulfides (22.5% < x < 29.5%) is shifted to a lower range of x in relation to corresponding selenides (28% < x < 37%), as shown in Figure 8. To put these results in perspective, we have assembled known IPs in group V sulfides and selenides [21]. The IP in P-Se and As-Se binary systems display close parallels; their centroids are nearly the same, although IP width of the P-Se binary ($\Delta r = 0.12$) exceeds that of the As-Se ($\Delta r = 0.08$) one. We believe that aspects of local structure, directly accessible from experiments, most likely play a role in determining these widths. There are two local structures, PYR ($r = 2.40$) and QT units ($r = 2.28$) in As-Se glasses that contribute to the width of the IP. On the other hand, there are three local structures that contribute to the IP width in P-Se glasses; PYR ($r = 2.40$), QT ($r = 2.28$) and a polymeric Ethylene-like, $P_4Se_4$ unit (ETY) ($r = 2.5$). The number appearing in parenthesis describes the mean coordination number of these units or chemical stoichiometry. While the PYR and QT units are isostatic, the ETY units are mildly stressed-rigid. The somewhat larger width of the IP in P-Se glasses than in As-Se ones, most likely, derives from the wider range in $r$ spanned by the three local structures. On the other hand, the similarity of IP centroids comes largely from the fact that both these group V selenides form polymerized network structures in which $Se_n$ chains are part of the network backbone (Figure 17).

Moving next to the cases of the binary As- and the P- sulfides, one is struck by the shift of IP centroids to lower x in relation to corresponding selenides (Figure 19). What aspect of glass structure contributes to the shift? It is instructive to first consider the case of P-S binary, where there is recognition [21] that the IP centroid shift to lower x results from a complete segregation of the $S_n$ chain fragments from the PYR and QT units bearing connective tissue (backbone). In Raman scattering, vibrational modes of QT, PYR and $S_n$ chains are completely resolved in binary P-S glasses. The concentrations of these local structures can then be estimated, and one finds that the chemical stoichiometry of the IP backbone equals $P_{33}S_{66}$ if all excess S is to completely segregate from the backbone. On the other hand, in P-Se glasses IP centroid stoichiometry, $[x_c(1) + x_c(2)]/2 = [29\% + 37\%]/2$ is equal to 33%, the <u>same</u> as found in P-S glasses. We are thus led to think that the IP centroid shift in P-S glasses is largely a manifestation of the excess S (as $S_n$ chain fragments) completely decoupling from the backbone.

A perusal of the data of Figure 19 shows that the IP centroid in binary As-S glasses is shifted halfway between the case of P-Se ($Se_n$ chains fully coupled to network ) and that of P-S ( $S_n$ chains fully decoupled from the network) glasses. Thus, a plausible explanation is that about half of the excess S forms part of the backbone while the other half is decoupled from the backbone in binary As-S glasses. The demixed excess S cannot be present in short (< 8 atom long) $S_n$ chain fragments, otherwise one would have seen end effects. The demixed $S_n$ chains are most likely at least 15 atoms or longer. To put the number in perspective, Kozhevnikov et al.[75] in their studies of the depolymerization transition in elemental sulfur place the length of sulfur chains close to 100 nm long or about 600 atoms long. The excess S in the IP is viewed as present in chains and not rings as suggested by Raman scattering (Figure 9).

What can we say about the width of the IPs in the sulfide glasses? For the companion binary P-S glasses, an IP is observed in the 16% < x < 19% range ($\Delta r$ = 0.03), and one finds QT units and PYR units concentrations to show global maxima respectively at the lower end ( x = 16%) and at the higher end (x = 19%) of that phase. We believe a parallel circumstance prevails in the As-S binary with both PYR and QT units contributing to the width of the IP. We have already noted that QT unit concentrations display a maximum in the IP (Figure 11). On the other hand, PYR unit concentration increases monotonically with x to maximize near x = 40% in the As-S binary (Figure 11). This particular feature of the data on As-S glasses differs from the case of the P-S glasses, in large part because P-P bonds ( 51.3 kcal/mole) have a higher chemical bond strength than As-As ( 32.1 kcal/mole) bonds, and these bonds readily are manifested in $P_4S_{10}$ molecules that decouple from the backbone just above the IP.[21].

Nanoscale phase separation effects, such as formation of small molecules that demix from a backbone, result in loss of network character , a structural feature that plays an important part in determining IP widths in glasses. We have already noted[21, 34] that segregation of $S_8$ rings at low x (< 14%) and of $P_4S_{10}$ molecules at high x (> 20%) serves to restrict the range of glass formation, and the IP as well. Parallel considerations must also apply to the case of the present $As_xS_{1-x}$ binary, wherein at x < 18% concentration of $S_8$ rings proliferate while at x > 38%, $As_4S_4$ molecules decouple [22] from the network. In fact bulk glass formation in the As-S binary ceases at x > 50%, largely because of the preponderance of $As_4S_4$ molecules in melts, which upon cooling give rise to the crystalline phase, Realgar[76]. IPs in network forming systems require that there be a significant fraction of the connective tissue present for a glass network to form an IP and self-organize.

The broad picture of partial demixing of $S_n$ chains in the IP of As-S glasses described above is corroborated by molar volumes and $T_g$ of glasses. Partial demixing of $S_n$ chains from network backbone contributes to the increased free volume of As-S glasses (Figure 4) as discussed earlier in section IV. The almost complete absence of chemical bond strength scaling of $T_g$ between As-Se and As-S glasses, and also in the IP (Figure 7), is suggestive that network connectivity of sulfide glasses is <u>lower</u> than of the counterpart selenides. In summary, the shift of the IP to lower x, the increased molar volumes and the complete absence of bond-strength rescaling of $T_g$s in the As-S glasses in relation to As-Se glasses, each point to the partial demixing of $S_n$ chain fragments from network backbone.

### C. Low frequency vibrational modes; evidence of boson and floppy modes

An almost universal feature of disordered solids is the appearance of vibrational modes at low frequency (5- 60 cm$^{-1}$ range) in Raman scattering, inelastic neutron scattering and low temperature specific heats broadly known as 'boson modes". These modes are identified with an excess of vibrations over Debye-like ones ($g(\upsilon) \sim \upsilon^2$), and probably come from some transfer of strength from high to low frequencies as an ordered crystalline solid is rendered disordered-amorphous. The nature of these excess vibrations has been a subject of ongoing debate over the past two decades or more[77-79]. Most work on boson modes has been performed on stoichiometric glasses and polymers, and in some cases the role of temperature and pressure[80] has also been investigated. In select cases compositional studies on glasses have also been performed[81]. Recently, attention has been paid to analysis of low frequency Raman modes to extract meaningful boson parameters from the excess vibrational density of states (e-VDOS) that could be related to other probes of these vibrations.

In the present work we have paid particular attention to analyzing the low frequency Raman scattering data in over a wide range of compositions (8 < x < 41%) encompassing the flexible, intermediate and the stressed rigid phases of $As_xS_{1-x}$ glasses. Details of the analysis were presented in section IV. Perusal of the data of figure 15b shows that the slope of the normalized integrated intensity $I_{edos}/I_{edos}(x = 0)$ with x yields a value of the slope, $d(I_{edos}/I_{edos}(x = 0))/dx = 1.6(1)$. The result is reminiscent of the floppy mode fraction change with $r$ in rigidity theory[16]. in In a 3D network of chains or rings, wherein each atom has 2 near neighbors, or $r = 2$, there is one floppy mode (f = 1) per atom. Upon alloying As with S, the resulting glass network becomes increasingly cross-linked as the mean coordination number, $r = 2 + x$ increases, and the floppy mode count f($r$) steadily decreases,

$$f = 6 - (5/2)r \qquad (5)$$

and as x approaches 40% , or $r = 2.40$ and networks becomes rigid. In such a mean-field description of the onset of rigidity, one finds that the slope, $df/dr = 2.50$ (equation 5). Given that $r = 2+ x$, the observed slope as a function of $r$, also represents the slope as a function of x, i.e., $d(I_{edos}/I_{edos}(x=0))/dr = d(I_{edos}/I_{edos}(x = 0))/dx = 1.6(1)$. In other words, the observed variation in the integrated intensity of the low-frequency modes as a function of $r$ in our experiments of 1.6 is somewhat lower than the rigidity theory prediction of the slope of 2.5.

The most natural interpretation of these data is that the low-frequency vibration modes in these glasses has contributions arising from two sources, (i) floppy modes associated with the 2-fold coordinated S atoms in rings and chains, and (ii) soft modes associated with a domain structure of glasses ( Figure 15b) . The latter modes have been discussed by Duval et al.[77], who have suggested that glasses may be visualized as composed of a domain structure, with domains representing nanometer sized elastically stiff regions in which atoms are strongly bonded and move together as a unit against other domains. The boson mode is viewed as an inter-domain soft acoustic vibration, and its frequency given by

$$\omega = v_s/2cR \qquad (6)$$

where $v_s$, c, and R represent respectively the velocity of sound, velocity of light and the mean size of a domain. Taking velocity of sound in pure S to be 1250 m/s[75] at 196°C and of $As_2S_3$ glass to be 2100 m/s[82], we have interpolated the speed of sound for intermediate compositions. The domain size, R, suggested from equation (5), is then found to increase monotonically from 0.55 nm near x = 5% to 0.75 nm near x = 45%. The trend of an increase in R with x appears quite plausible given that binary glasses become more connected, although it is at present less clear if the magnitude of R can be identified with characteristic clusters in these glasses.

The reversibility window in the present glasses shows that the rigidity transition in the present glasses occurs near x = 22% (window onset composition $x_c(1)$) and the stress transition near 30% (window end composition, $x_c(2)$). Within a mean-field description one expects the floppy mode contribution f(x) to nearly vanish as x increases to 22%. Thus, a possible description of the low-frequency vibrations in these glasses is that (Figure 15(b)) these are dominated by floppy modes at low x, and soft modes at high x. The soft or boson mode contribution most likely comes from the intrinsically heterogeneous character of the present sulfide glasses resulting from the partially polymerized nature of the backbone structure. Structural groupings such as $S_8$ rings, $S_n$ chain fragments that decouple from the network backbone will contribute to the soft mode behavior.

The onset of the flexible phase at x < 22.5% signaled by the calorimetric measurements, and the growth in low frequency vibrations in Raman scattering at x < 20%, is a striking result. These findings illustrate that two completely independent probes, light scattering and calorimetry, come together in quantitatively probing the flexible phase of the present glasses in a rather convincing fashion. The Raman and calorimetric data on the present gasses, for the first time, quantitatively connects boson modes and floppy modes to the low frequency vibrations observed in a network glass.

CONCLUSIONS

Dry and homogeneous bulk $As_xS_{1-x}$ glasses, synthesized over the 8% < x < 41% range, were investigated in modulated DSC, Raman scattering, Infrared reflectance and molar volume measurements. The following conclusions emerge from this work. (a) Vibrational mode assignments, assisted by first principles cluster calculations, have permitted decoding aspects of local structures, and show that both pyramidal $As(S_{1/2})_3$ and quasi-tetrahedral $S=As(S_{1/2})_3$ local structures are formed in these glasses. (b) Modulated DSC experiments reveal existence of a reversibility window in the 22.5% < x < 29.5% range that is identified with Intermediate Phase (IP), with glasses at x < 22.5% in the *flexible* phase and those at x > 29.5% in the *stressed-rigid* phase of the present As-sulfides. (c) Molar volumes of the glasses decrease almost linearly with increasing x displaying a local minimum in the IP. (d) The shift of the IP centroid in the present

As-sulfides to lower As concentration x compared to the IP in corresponding As-selenides is identified with nearly half of the excess sulfur decoupling from the backbone in the present sulfides in contrast to all excess Se forming part of the network backbone in corresponding As-selenides. (e) Presence of both pyramidal $As(S_{1/2})_3$ and quasi-tetrahedral $S=As(S_{1/2})_3$ local structures as building blocks of these glasses provide the structural variability to understand, qualitatively, width of the IP, $\Delta x = 7.0\%$ in glass composition or $\Delta r = 0.07$ in mean coordination space with the QT units contributing to the lower end and PYR units to the upper end of the window. (f) Low-frequency Raman modes, generally identified as boson modes, increase in scattering strength linearly as As content x of glasses decreases from x = 20% to 8%, with a slope that is close to the floppy mode fraction in flexible glasses predicted by rigidity theory. The result shows that floppy modes contribute to boson modes in flexible glasses. In the intermediate and stressed rigid elastic phases, scattering strength of boson modes persist at a constant level, a finding we attribute to the presence of soft modes associated with domain structure of glasses. Both floppy modes and soft modes both contribute to boson modes in the present glasses.

Acknowledgements

It is a pleasure to acknowledge discussions with Professor Bernard Goodman, Professor Eugene Duval and Professor Darl McDaniel during the course of this work. This work is supported by NSF grant DMR 04- 56472 to University of Cincinnati.

**Figure captions**

**Figure 1**. (Color online). Glass transition temperature in $Ge_xP_xS_{1-2x}$ (ref. 42) and $Ge_xP_xSe_{1-2x}$ (ref. 43) ternaries compared. In the 10% < x < 18% range one observes scaling of $T_g$ with chemical bond strengths. At low x (< 10%), $S_8$ rings steadily demix, sulfur polymerization transition $T_\lambda$ is manifested, and $T_g$s steadily decline in the sulfide glasses.

**Figure 2.** (Color online). Eigenvectors and eigen-frequencies of pyramidal (left column) and quasi-tetrahedral (right column) local structures obtained from first principles cluster calculations. Dark (blue) and light (yellow) atoms represent arsenic and sulfur respectively. Mode frequencies and Raman cross sections ($\sigma$) are indicated with each vibrational mode. See Table 1 for a summary of results.

**Figure 3**. Phase diagram of $As_xS_{1-x}$ binary taken from the work of Blachnik et al. refs. 49, 50. There is a eutectic near x ~1% but none in the region 40% < x < 1% of As where the reversibility window is observed in the present work.

**Figure 4**. (Color online). Variations in molar volumes of $As_xS_{1-x}$ (● red, present work) and $As_xSe_{1-x}$ glasses (■, blue) taken from the work of Georgiev et al. ref. 19 ). Molar volumes of crystalline $As_2S_3$ (○, ref. [83]) is found to be 15% lower than that c-$As_2Se_3$ (□, ref. [84]). The dashed curve represents 15% scaled down trend of $As_xSe_{1-x}$ molar volume data, and represents approximately the predicted molar volumes of $As_xS_{1-x}$ glasses normalized for chalcogenide atom size. Note the observed molar volumes are significantly higher than the predicted ones, suggesting $As_xS_{1-x}$ glasses to have much free volume, most likely coming from presence of non-bonding van der Walls interactions.

**Figure 5**. (Color online). m-DSC scans of (a) orthorhombic sulfur (b) $As_8S_{92}$ glass (c) $As_{15}S_{85}$ glass. In (a) the three endothermic events include solid-solid phase transformation ($T_{\alpha \to \beta}$), melting transition ($T_m$) and sulfur polymerization transition ($T_\lambda$) are observed. In (b) and (c), the total (green), reversing (blue) and non-reversing (red) heat flows are shown. Notice that the non-reversing heat flow associated with the $T_\lambda$ transition displays an exotherm followed by an endotherm in the glass samples but not in elemental sulfur. See text.

**Figure 6**. (Color online). A summary of the (a) reversing heat flow (blue) scans (b) non-reversing heat flow (red) scans at indicated As concentrations x in the As-S binary. Note that at x > 23%, $S_8$ ring concentration nearly vanishes, and only the glass transition endotherm is observed.

**Figure 7.** (Color online). Variation in $T_g$ (x) in the present $As_xS_{1-x}$ binary (red) compared to the one in the $As_xSe_{1-x}$ binary (blue) (taken from Georgiev et al. ref. 19). Note the absence of chemical bond strength scaling of $T_g$s in the 25% < x < 40% range. The broken curve (·······) gives the expected $T_g$s of the $As_xS_{1-x}$ glasses if $S_8$ rings had not decoupled from the glasses, and bond-strength scaling had prevailed. The extrapolated $T_g(x=0)$ for a $S_8$ ring glass and a $S_n$ chain glass are estimated at -50°C and about 95°C.

**Figure 8**. (Color online). Reversibility windows in binary As-S (●, red) from the present work, compared to the window in binary As-Se (▲, blue) taken from the work of Georgiev et al. ref. 19 . Notice the window centroid in the sulfide glasses is shifted to lower x in relation to the one in selenide glasses.

**Figure 9.** (Color online). FT- Raman scattering data on present As-S glasses at indicated glass compositions (x) on the right of each curve. Similar results have been reported by several other groups, such as ref. [8, 11, 59, 60]. Near x = 40%, features of $As_4S_4$ molecules ( labeled R for Realgar) segregate from bulk glasses as shown in the left inset. Furthermore, a mode near 500 $cm^{-1}$, found to be present even at x = 41%, is shown in the right inset. $PYR^{ss}$ and $PYR^{as}$ represent the symmetric stretch and asymmetric stretch of pyramidal units. $QT^{ss}$ and $QT^{as}$ represent the symmetric stretch and asymmetric stretch of quasi-tetrahedral units. $QT^{As=S}$ represents the stretch of the As=S double bond of QT units.

**Figure 10.** (Color online). Lineshape analysis of Raman scattering in binary As-S glasses at indicated compositions, x = 8%, 10% and 15%. The mode assignments are indicated in the middle panel. $S_n$: Sulfur chain mode; $S_8$: sulfur ring modes. Also see fig.9 and Table 1 for mode identification.

**Figure 11**. (Color online). Variations of Raman mode normalized scattering strengths ($A_n/A_{total}$) as a function of glass compositions x in the present $As_xS_{1-x}$ binary: Sulfur chain and ring modes (■, green) at 467 and 473 $cm^{-1}$, QT mode (○) at 333 $cm^{-1}$, QT mode (▲, blue) at 490 $cm^{-1}$, PYR mode (●, red) at 365 $cm^{-1}$. Vibrational mode strengths were normalized to the total area under various modes. Note that the two QT modes (○,▲) show a broad global maximum in the IP.

**Figure 12.** (Color online). Compositional trends in Raman mode frequency of characteristic vibrational modes of QT and PYR units showing a red-shift with increasing x.

**Figure 13**. Dispersive Raman spectra of $As_xS_{1-x}$ glasses showing the boson mode at low frequencies at indicated compositions x on the right in percent: (a) observed Raman spectra ( $I_{expt}$) (b) Reduced Raman spectra ( $I_{expt}/[n_B + 1]$). Here $n_B$ is the Bose occupation number. Note the boson mode is the most important feature of the spectra in (a). In the reduced Raman spectra (b) the skewed lineshape at low frequency due to the finite T is removed, and the peak shifts up in frequency.

**Figure 14.** (Color online). Deconvolution of the reduced Raman spectra (red) ( $g(\omega)C(\omega)/\omega$) at two compositions, (a) x = 10% (top panel) and (b) x = 40% (bottom panel), showing extraction of the excess vibrational density of states (blue), e-VDOS, by subtracting a Debye like term (orange) from the total scattering.

**Figure 15**. (Color online). Variations in (a) the boson mode frequency (■) and (b) boson mode integrated intensity (■) as a function of glass composition x in $As_xS_{1-x}$ binary. In (b), the ---- line gives the floppy mode variation with $r = 2 + x$, and the ••••• line gives the difference between the integrated intensity (■) and the floppy mode (----) contribution.

**Figure 16**. (Color online). (a) Observed IR reflectance and (b) the imaginary part of dielectric constant $\varepsilon_2$ (TO response) in binary $As_xS_{1-x}$ glasses. The TO response was deduced from Kramers-Kronig transformation of the reflectance data.

**Figure 17**. (Color online). Infrared transverse optic TO (blue) and longitudinal optic LO (red)

response at select $As_xS_{1-x}$ glass compositions, (a) x = 40%, (b) x = 30% and (c) x = 20% compared. See text for details.

**Figure 18**. (Color online) Compositional trends in (a) $T_g(x)$(●, red) , $T_λ(x)$(Δ, blue) and the sulfur ring Raman mode near 217 cm-1( ▲, red) , (b) non-reversing enthalpy $ΔH_{nr}(x)$(●, red), and (c) molar volumes (■, blue) of binary As-S glasses from the present work. The reversibility window is observed in the 22.5% < x < 29.5% range, an observation that fixes the Intermediate Phase in the present sulfides. Also note that molar volumes show a local minimum in the IP.

**Figure 19**. (Color online). Observed IPs in group V chalcogenides. The bar chart gives the mean coordination number interval, Δr, across which IPs extend as noted by reversibility windows in m-DSC experiments.

Table 1. Raman and IR vibrational modes of As-S clusters. The frequencies and Raman/IR cross sections and depolarization ratios for each cluster are given, along with a description of the eigenvectors.

| Cluster | $\omega$(cm$^{-1}$) | IR | $I^{Ram}$ iso | $I^{Ram}$ total | $\rho$ | Description |
|---|---|---|---|---|---|---|
| AsS$_3$H$_3$ | 165 | 0.18 | 5.0 | 7.28 | 0.23 | umbrella |
|  | 352 | 0.18 | 31.3 | 34.0 | 0.06 | symmetric stretch |
|  | 355 | 0.78 | 0.00 | 6.63 | 0.75 | asymmetric stretch |
| AsS$_4$H$_3$ | 146 | 0.20 | 1.10 | 4.48 | 0.57 | umbrella |
|  | 335 | 0.21 | 59.5 | 60.7 | 0.01 | symmetric stretch |
|  | 365 | 1.17 | 0.00 | 14.3 | 0.75 | asymmetric stretch |
|  | 537 | 1.58 | 12.1 | 21.9 | 0.34 | S=As |

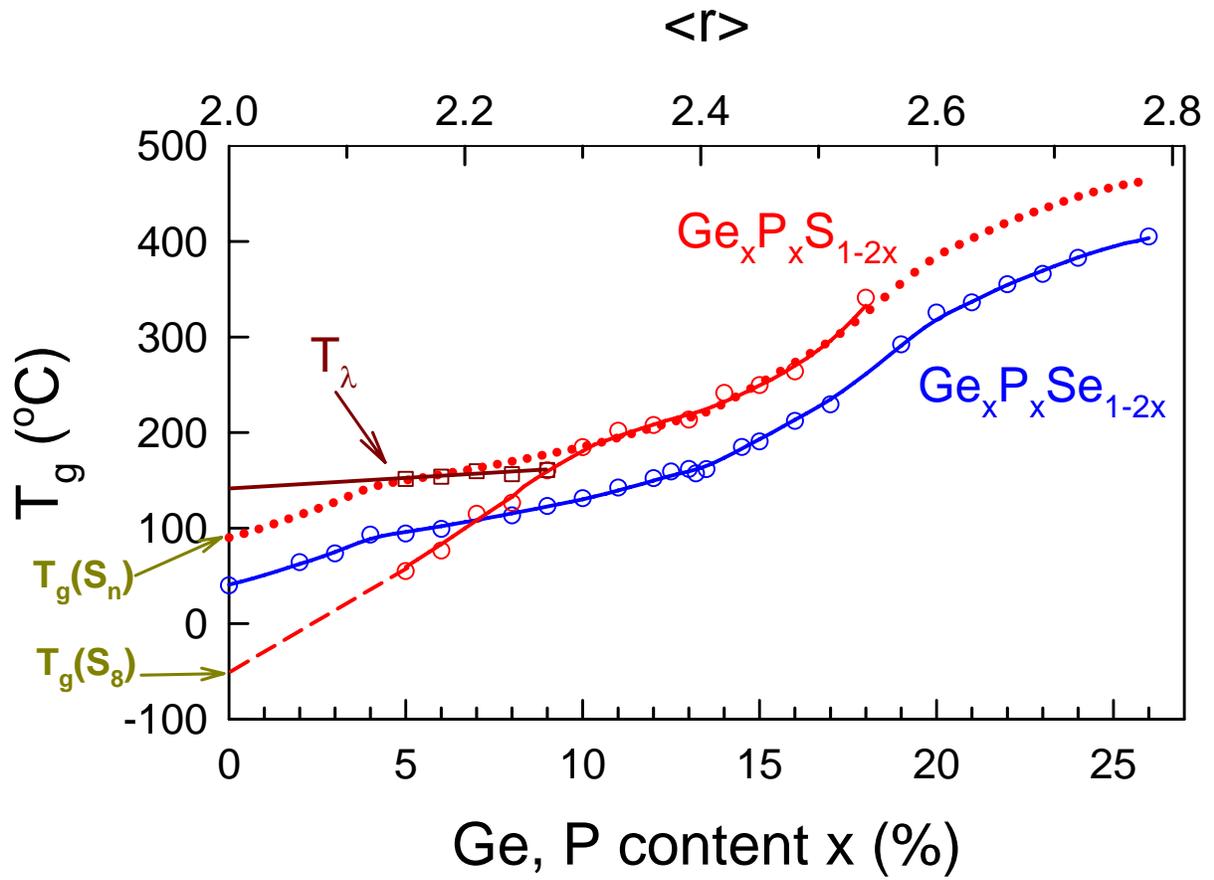

**Figure 1**

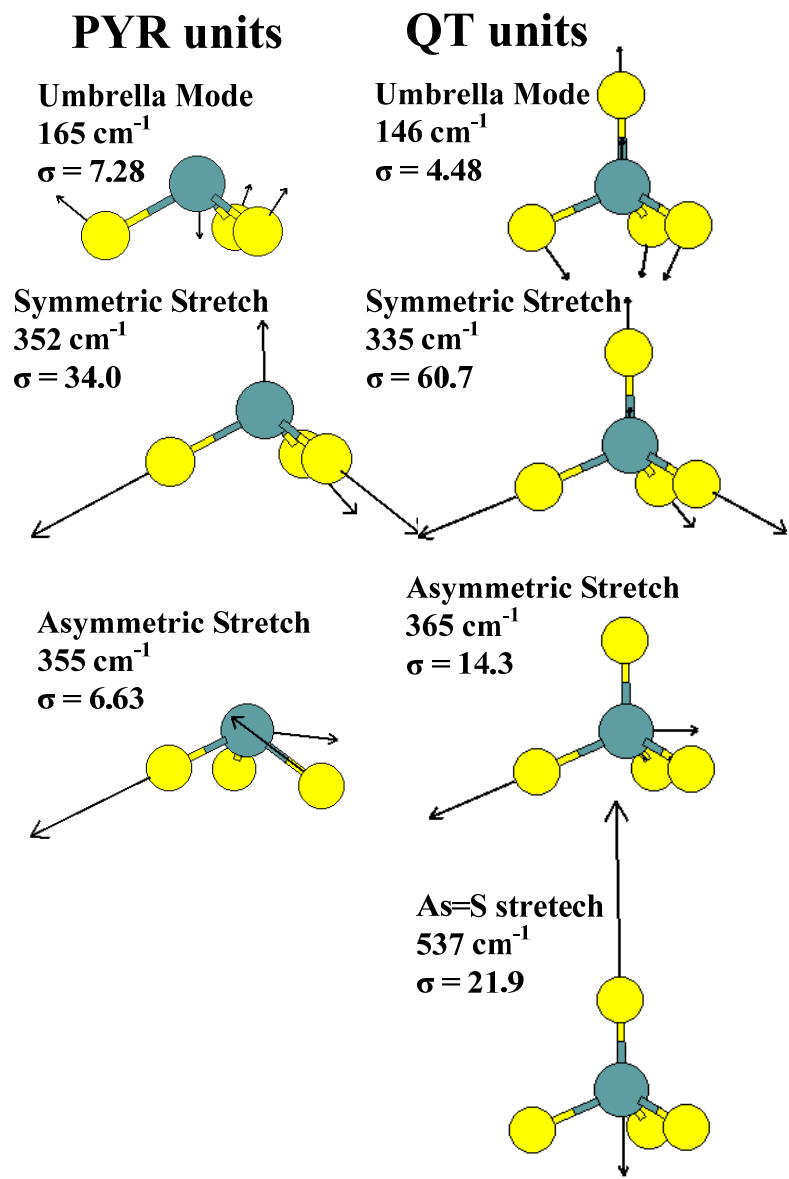

**Figure 2**

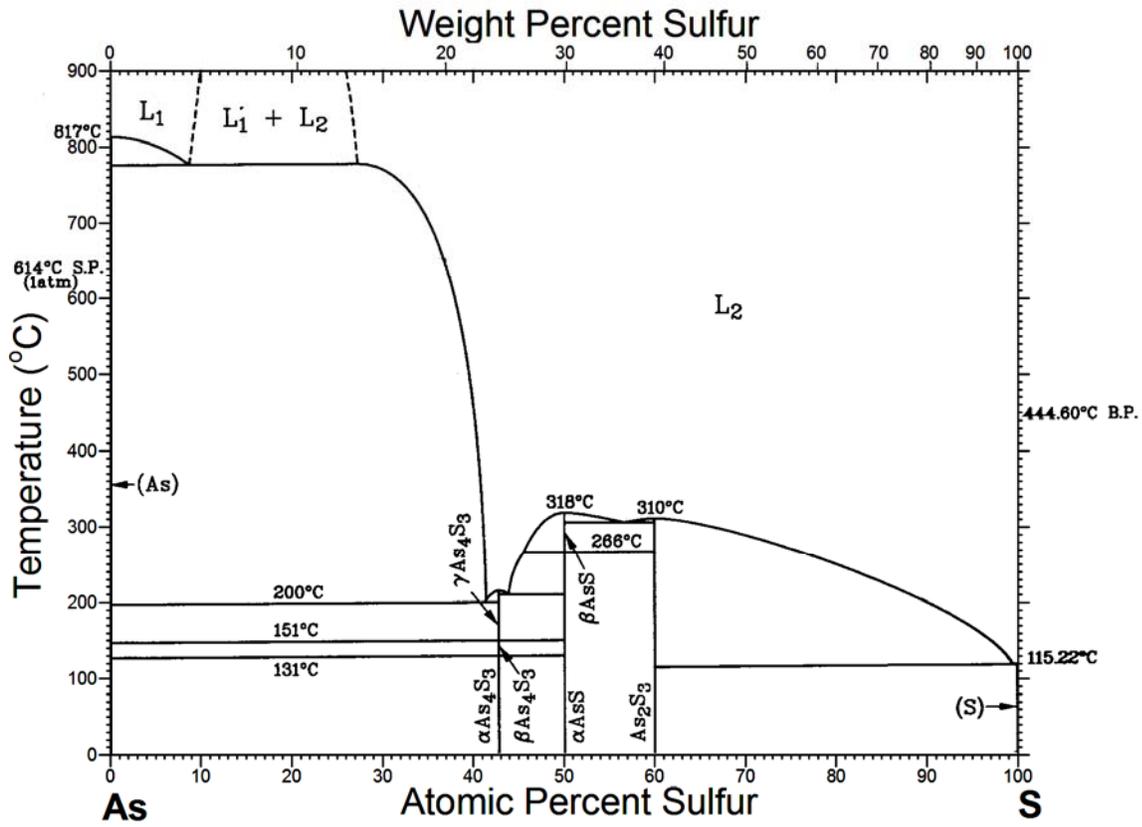

**Figure 3**

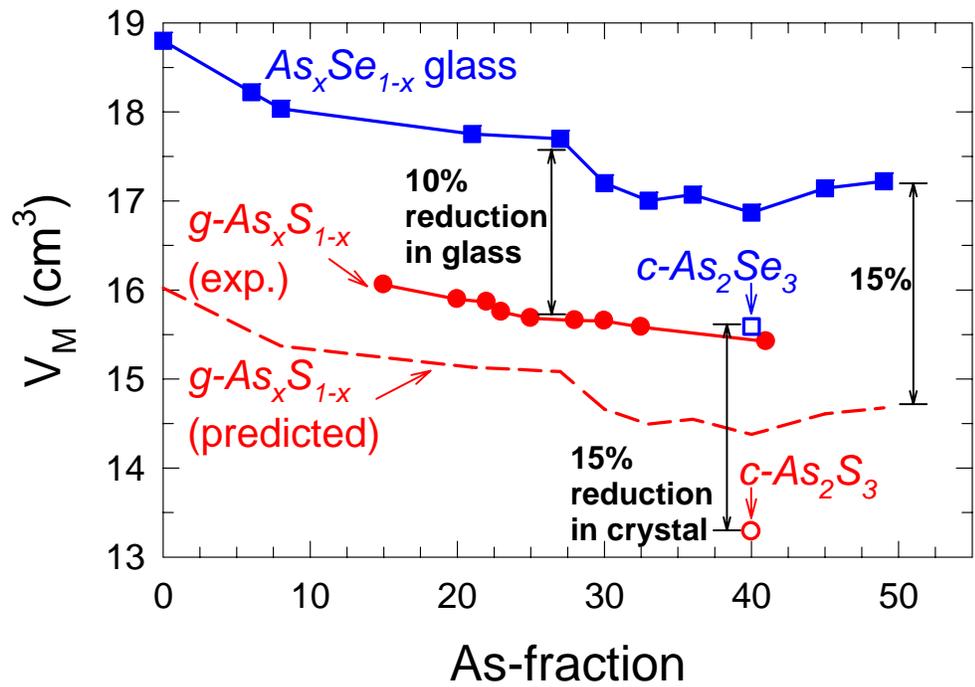

**Figure 4**

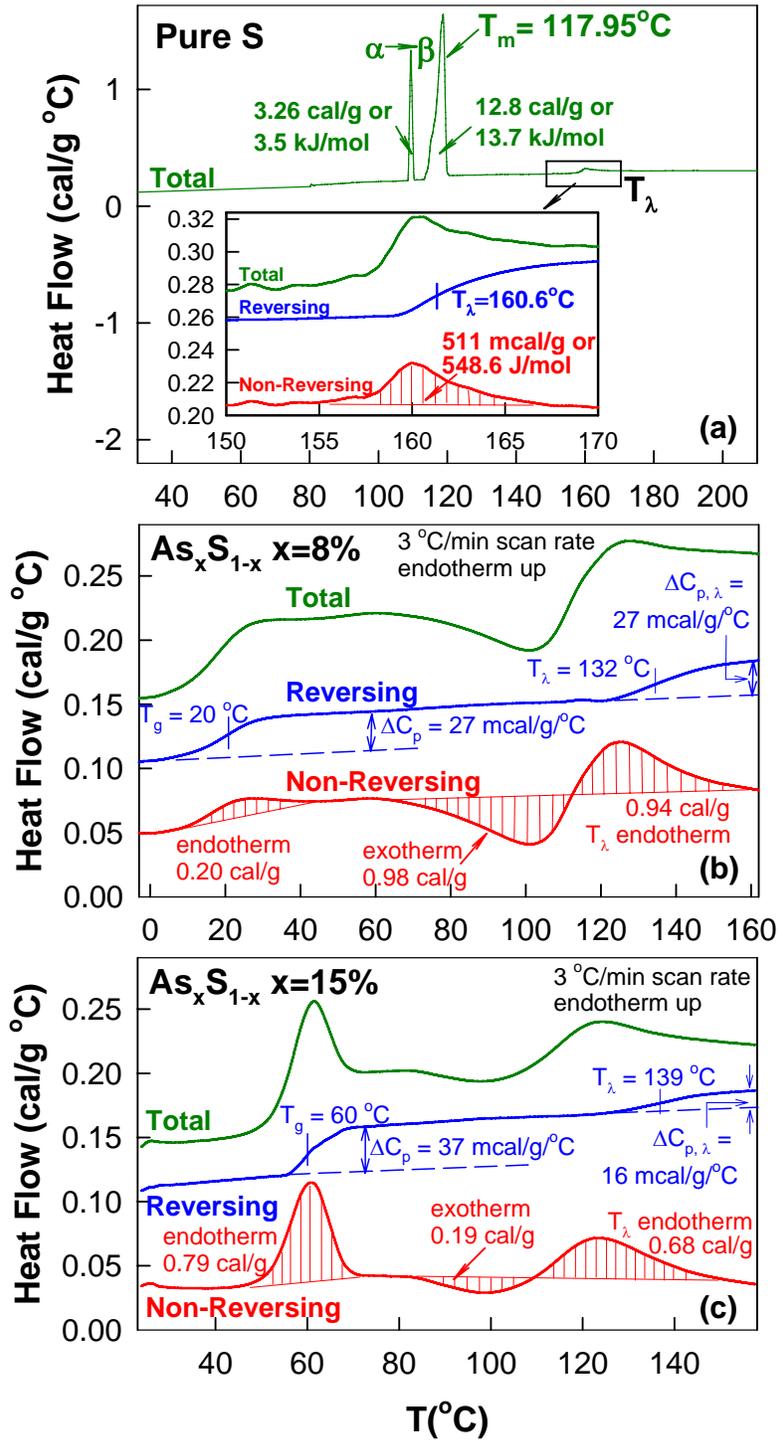

**Figure 5**

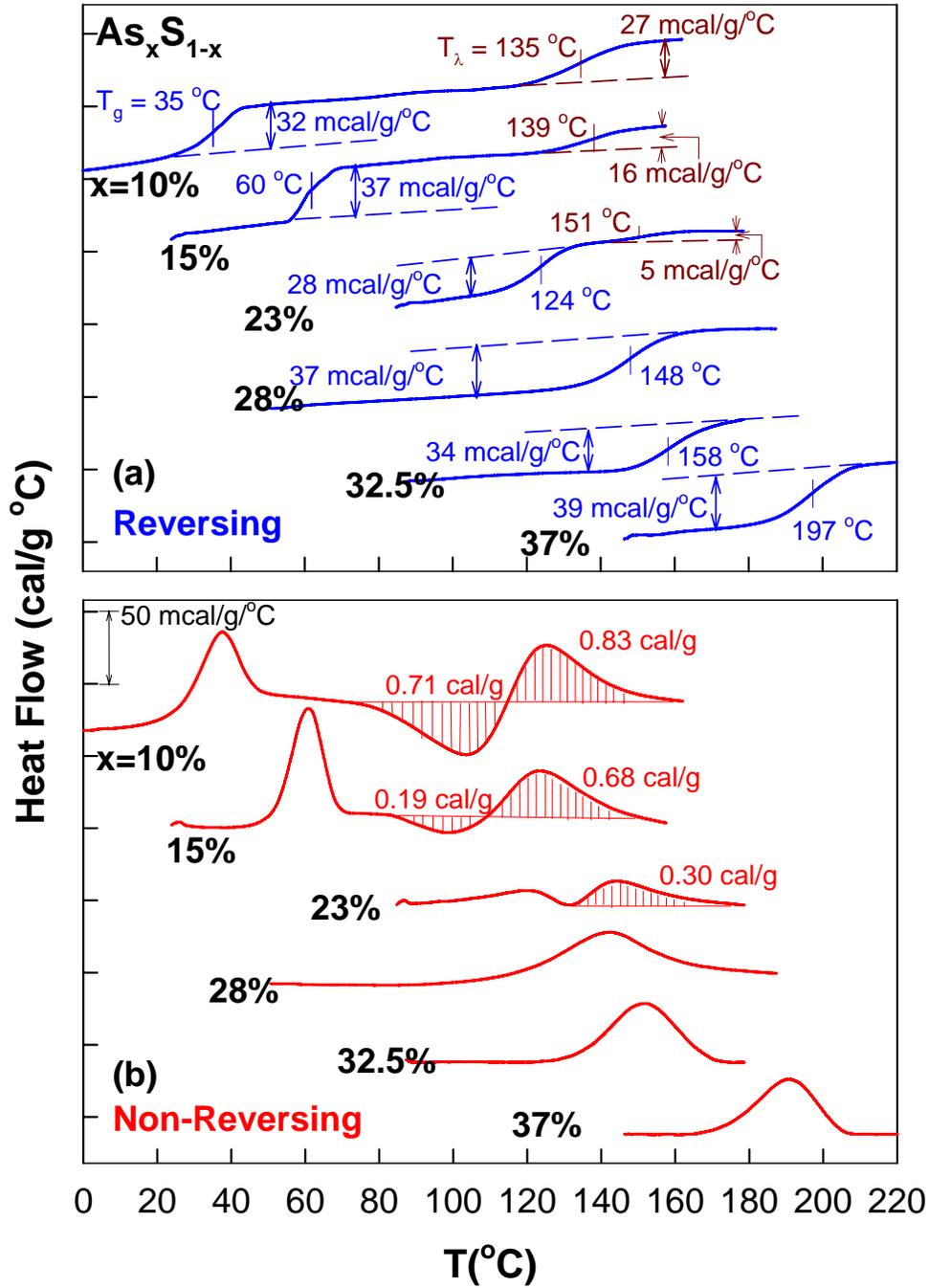

**Figure 6**

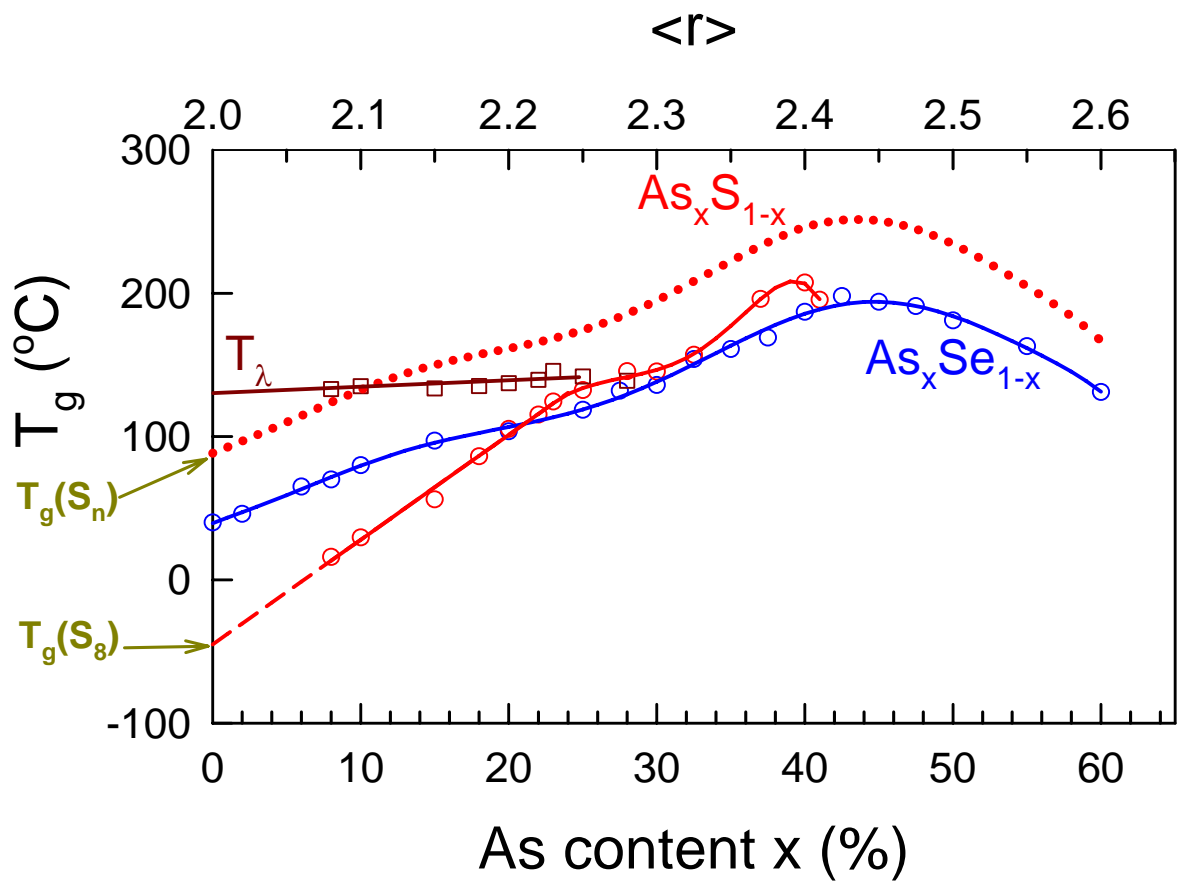

**Figure 7**

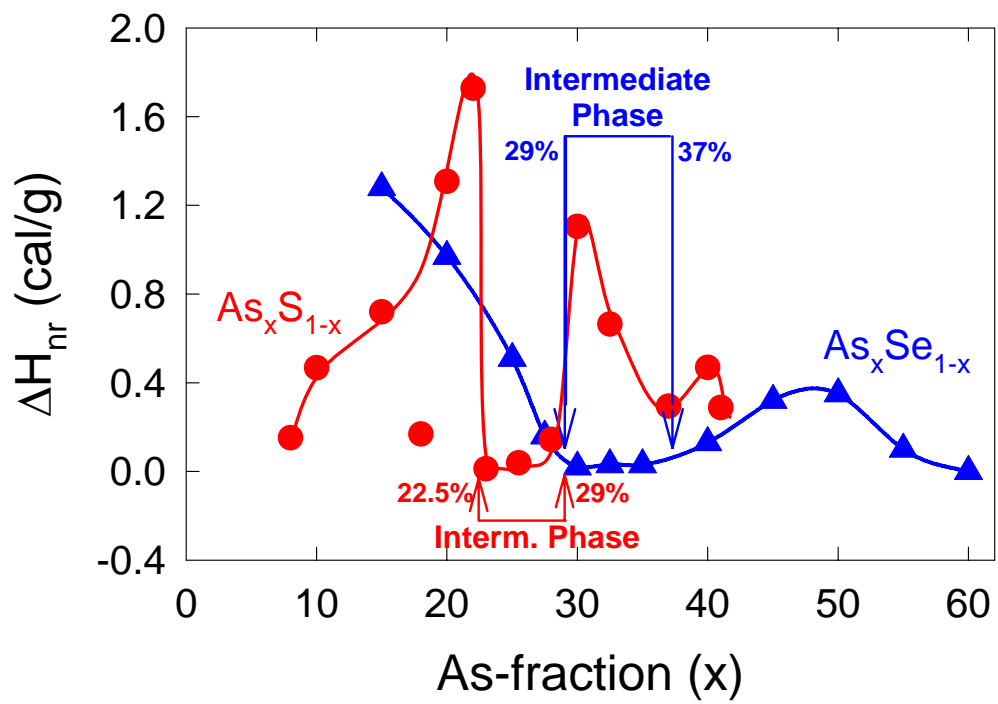

**Figure 8**

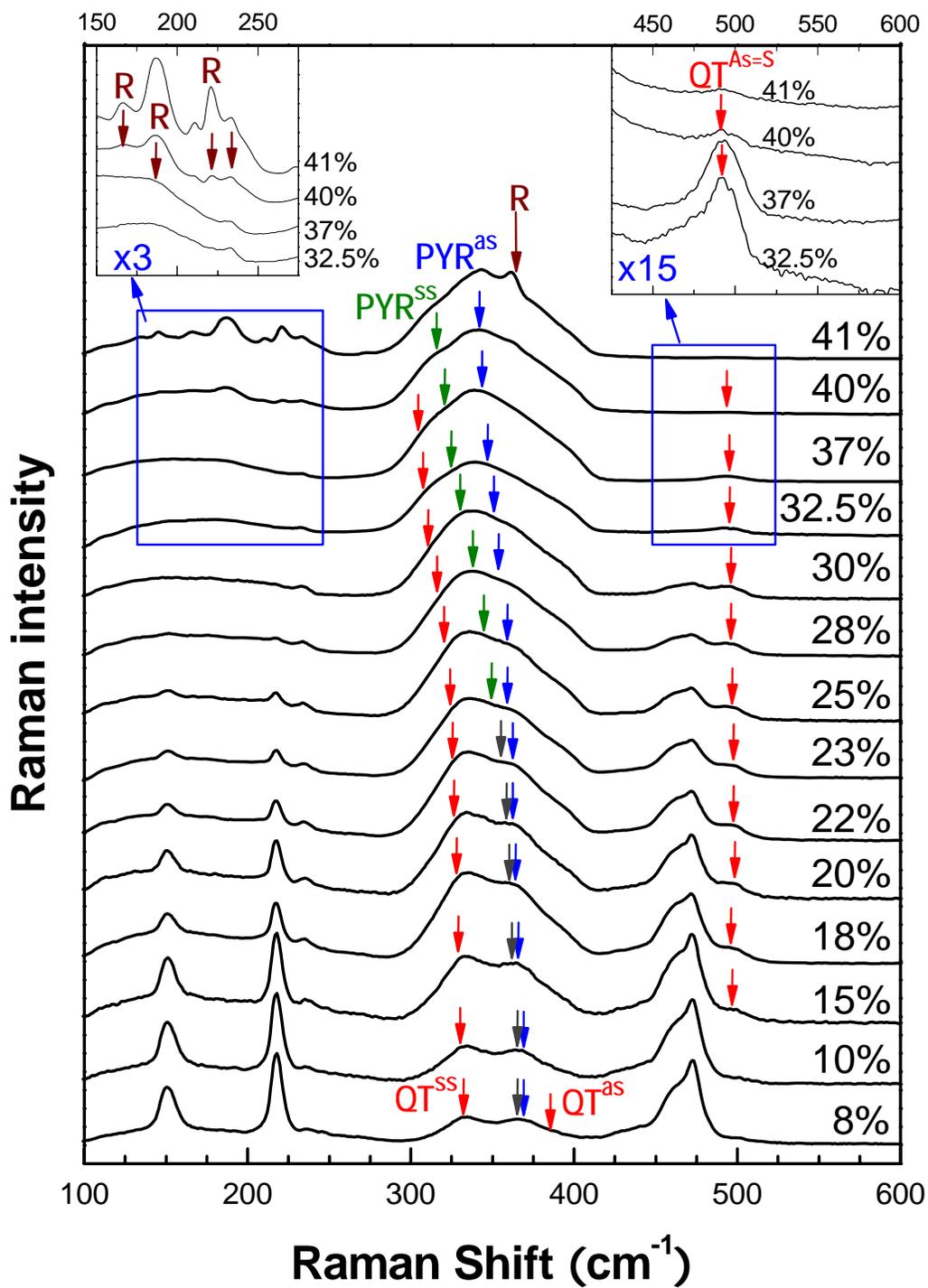

**Figure 9**

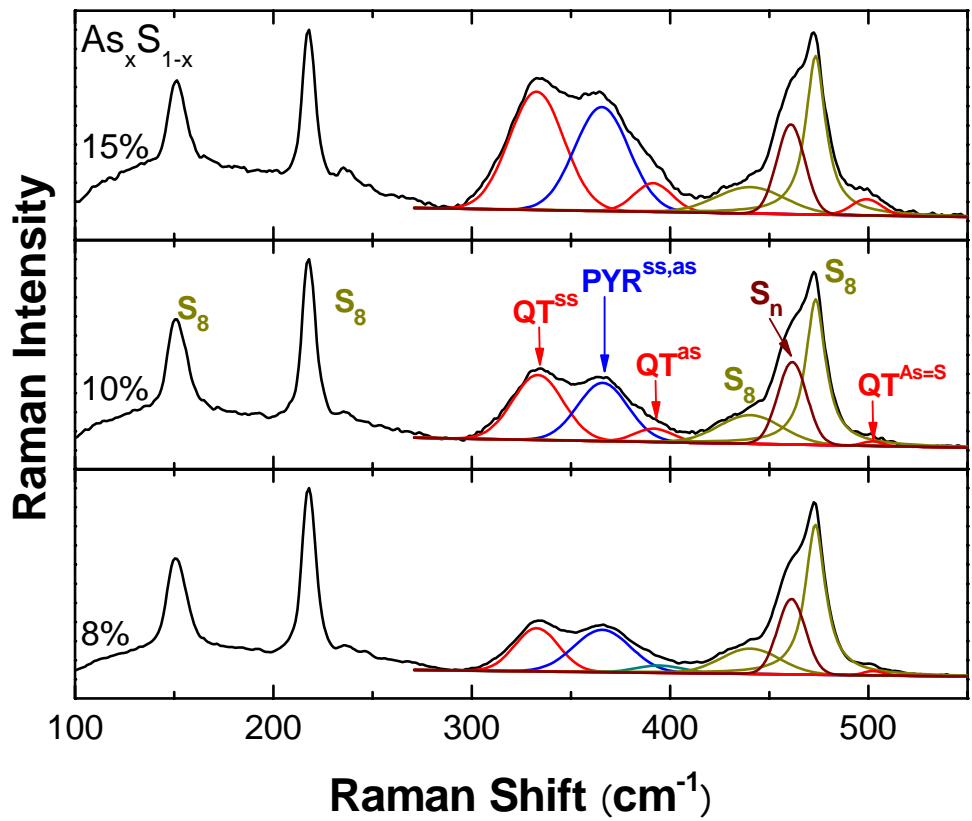

**Figure 10**

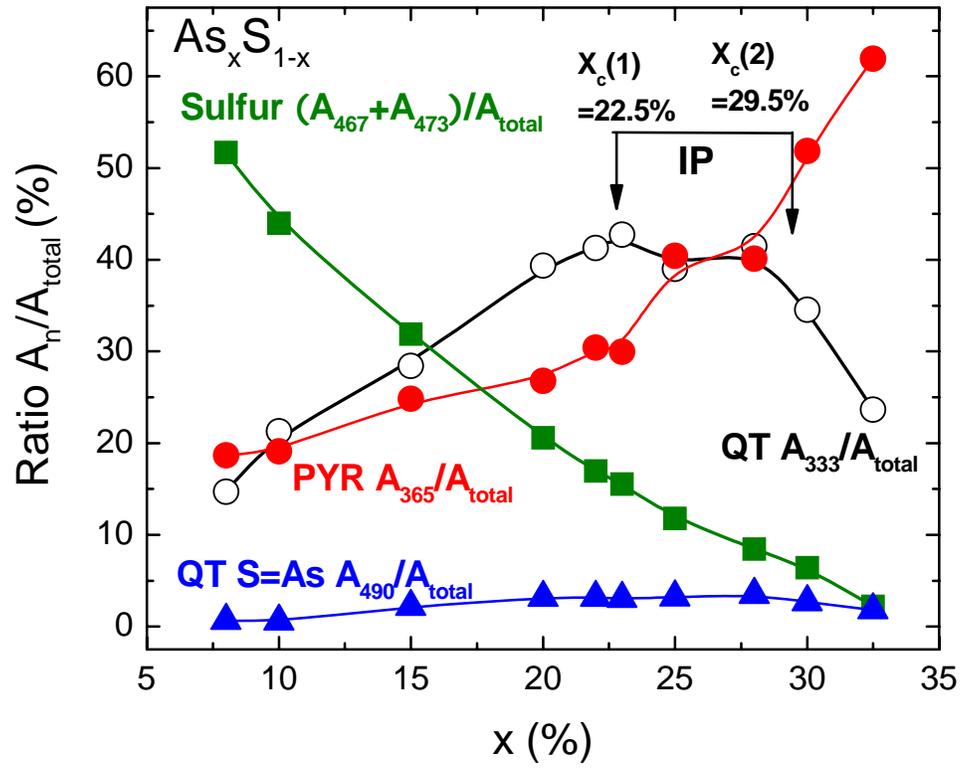

Figure 11

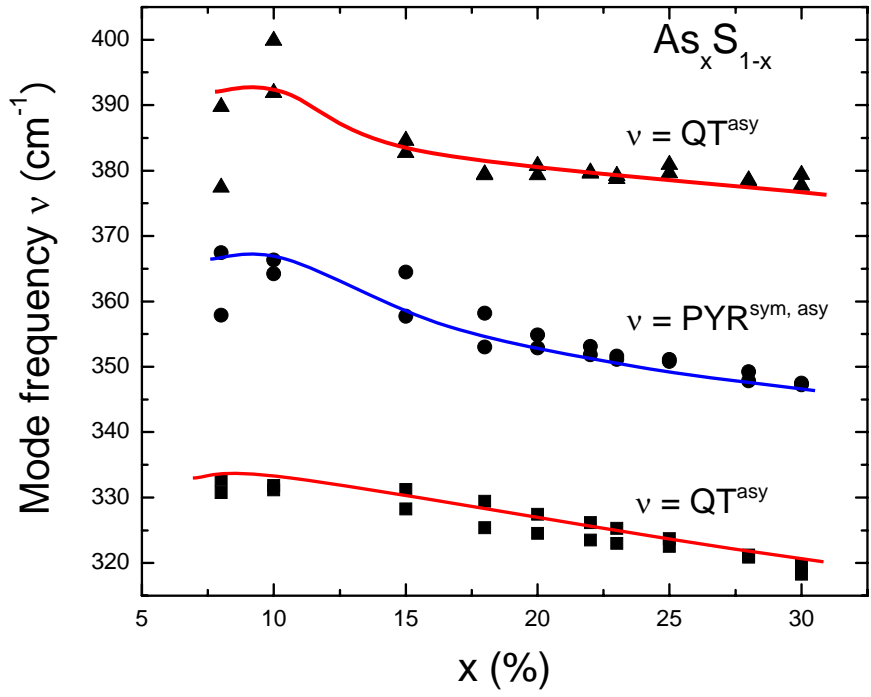

**Figure 12**

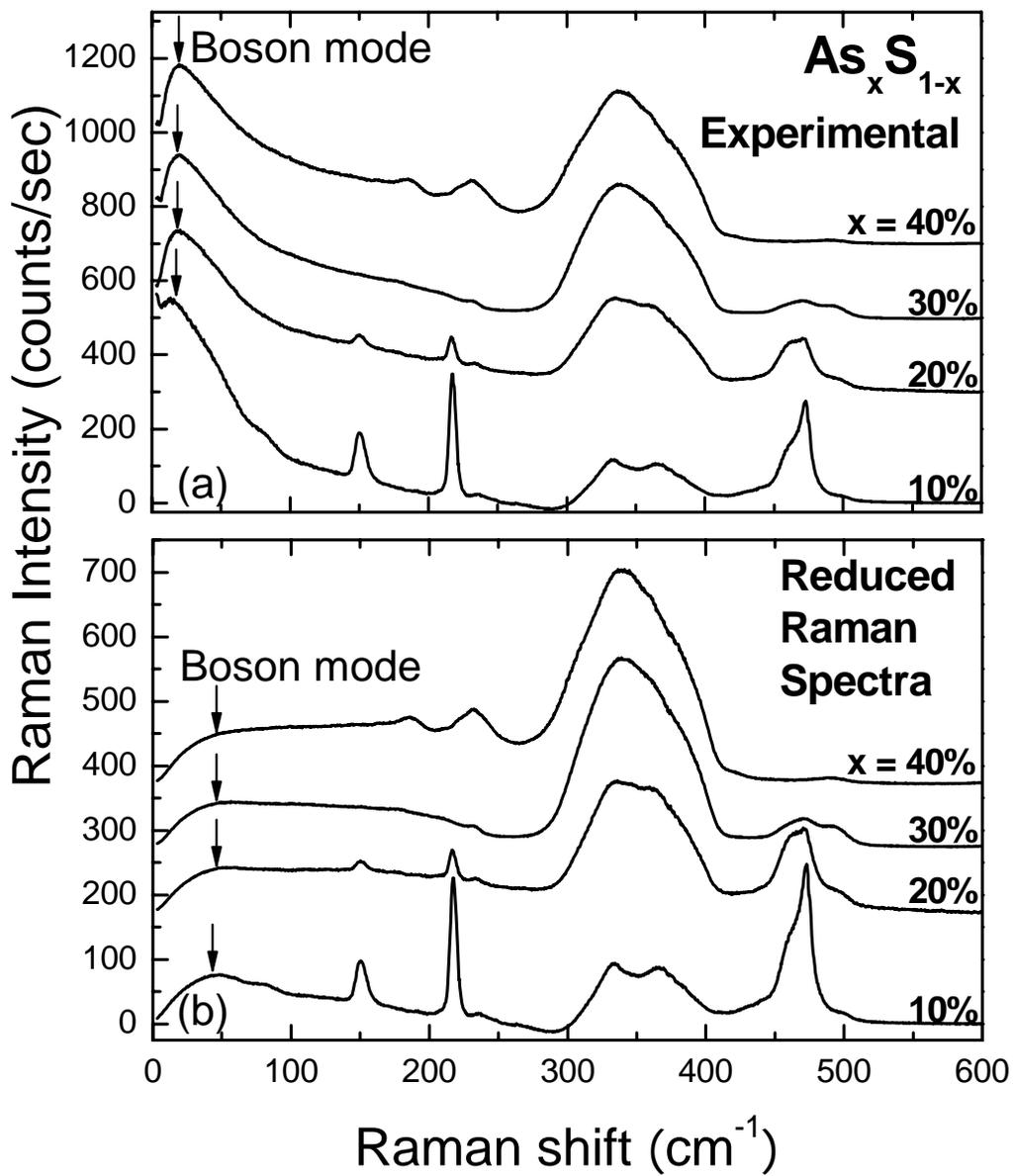

**Figure 13**

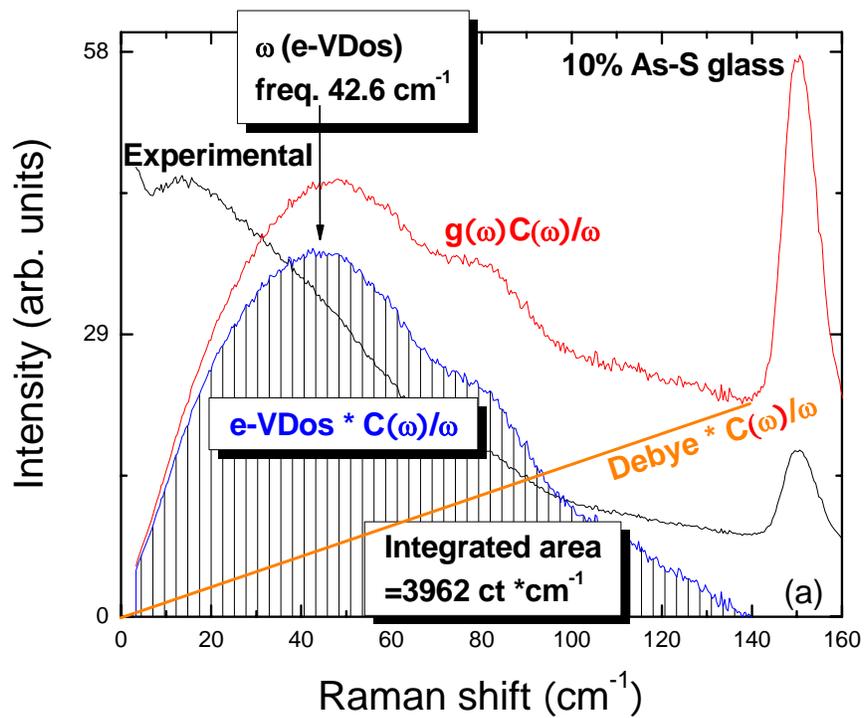
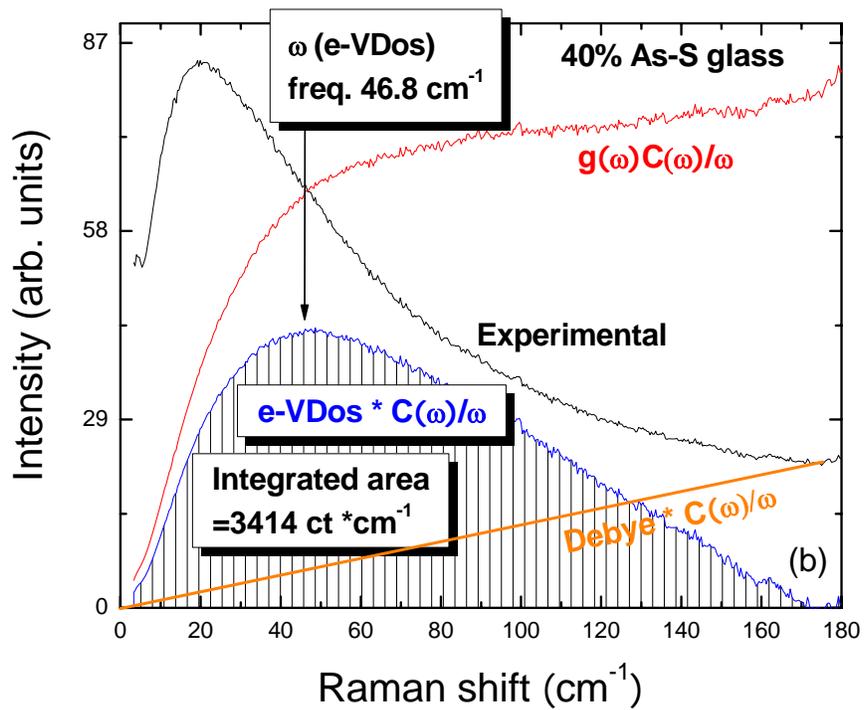

**Figure 14**

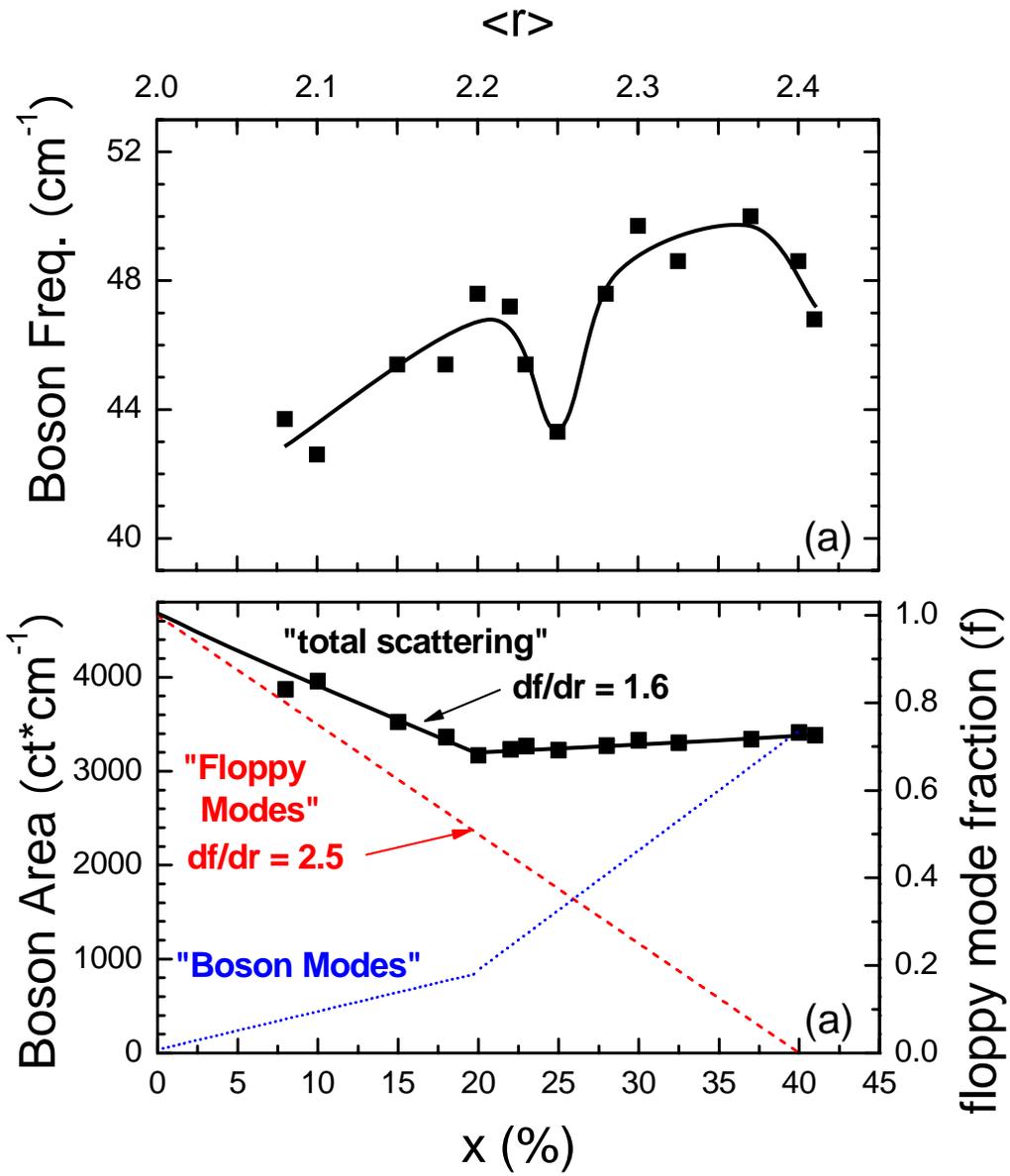

**Figure 15**

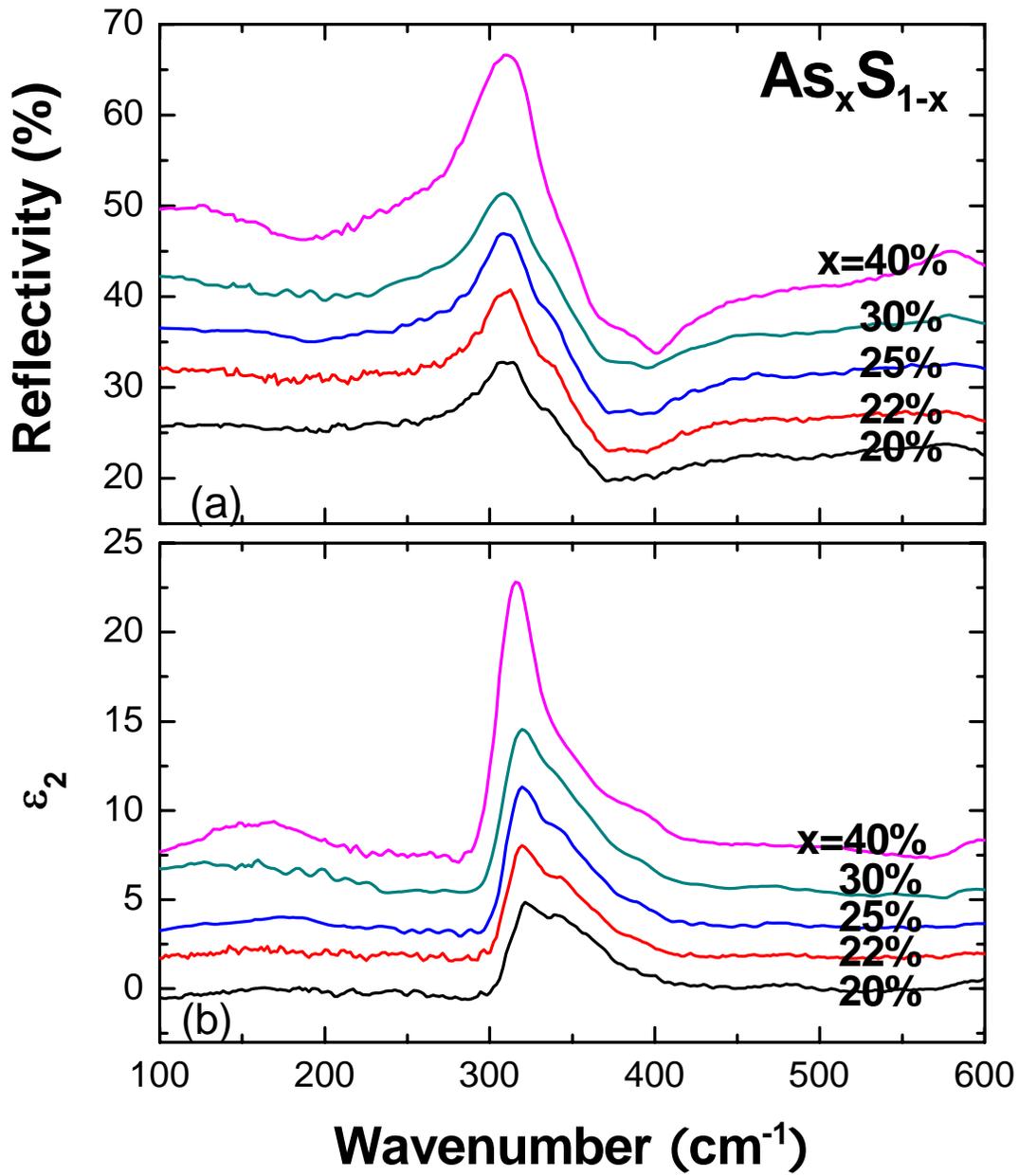

**Figure 16**

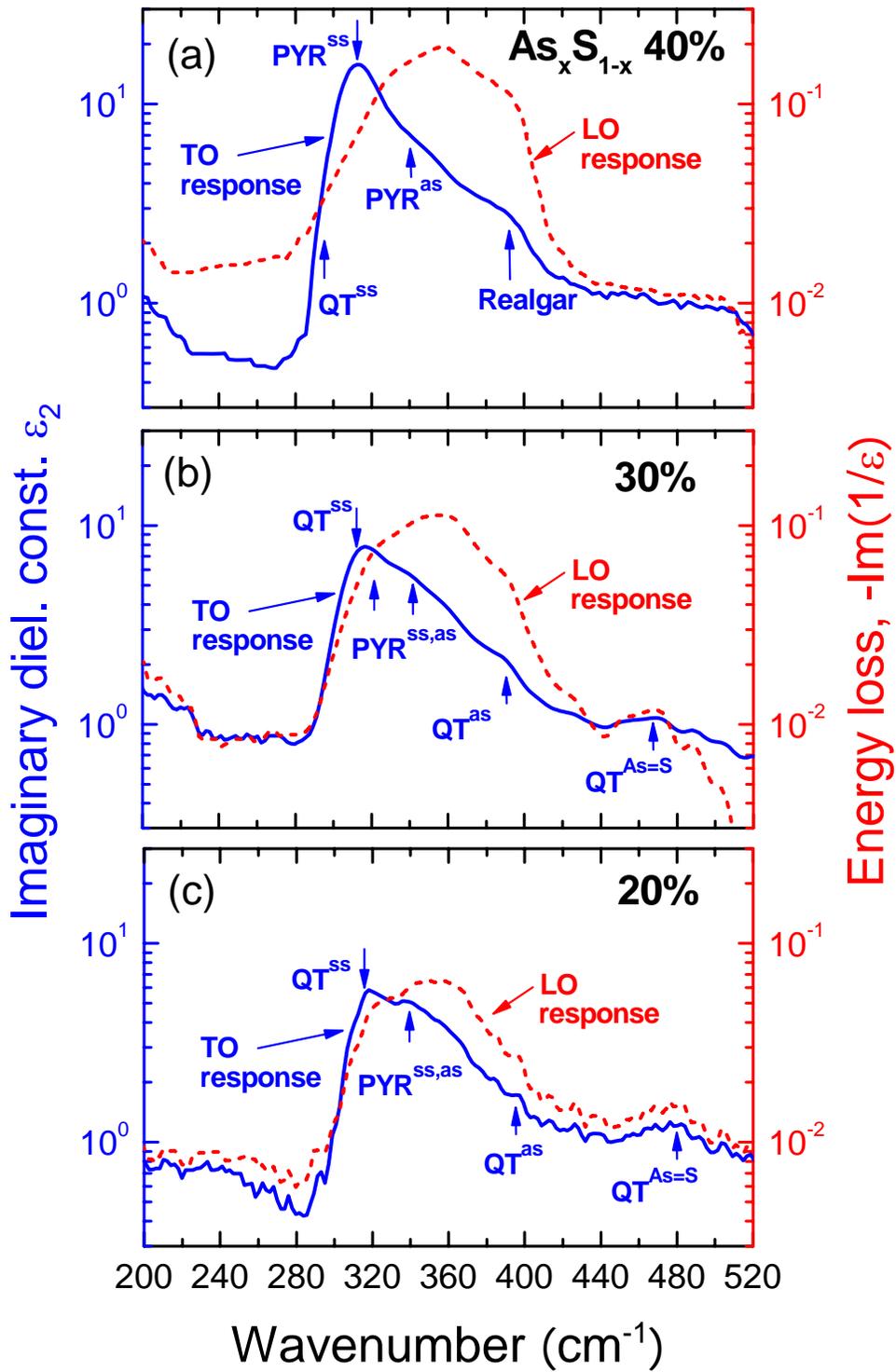

Figure 17

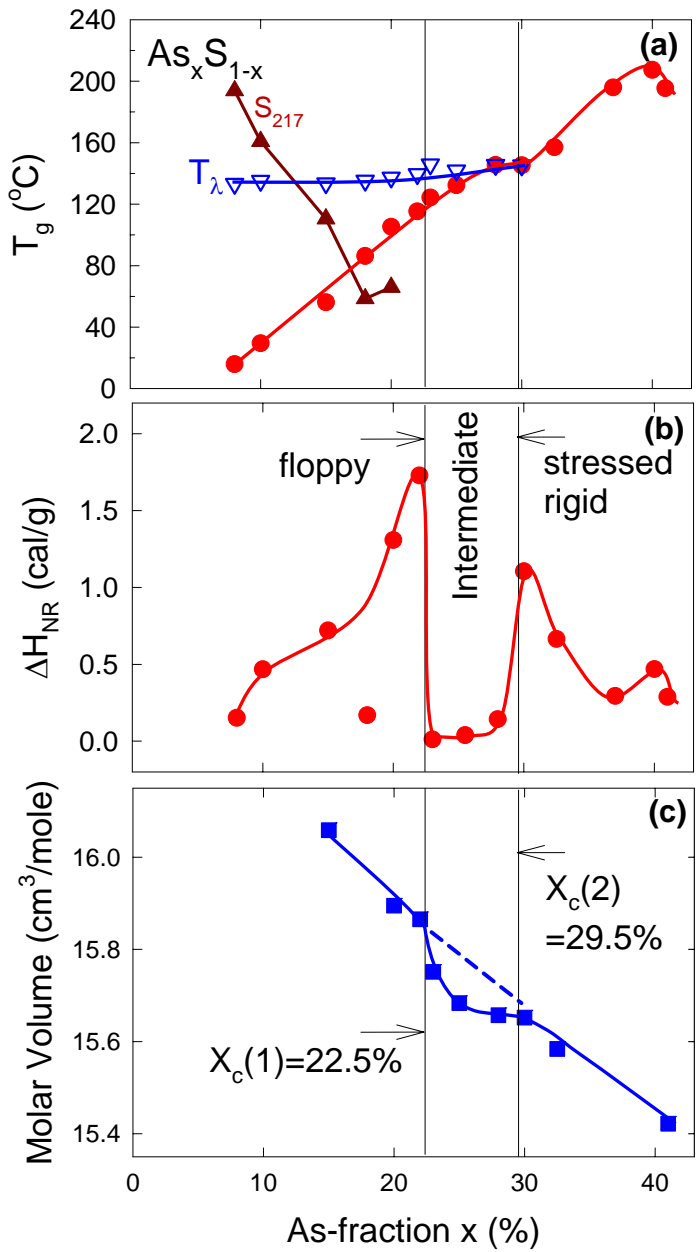

**Figure 18**

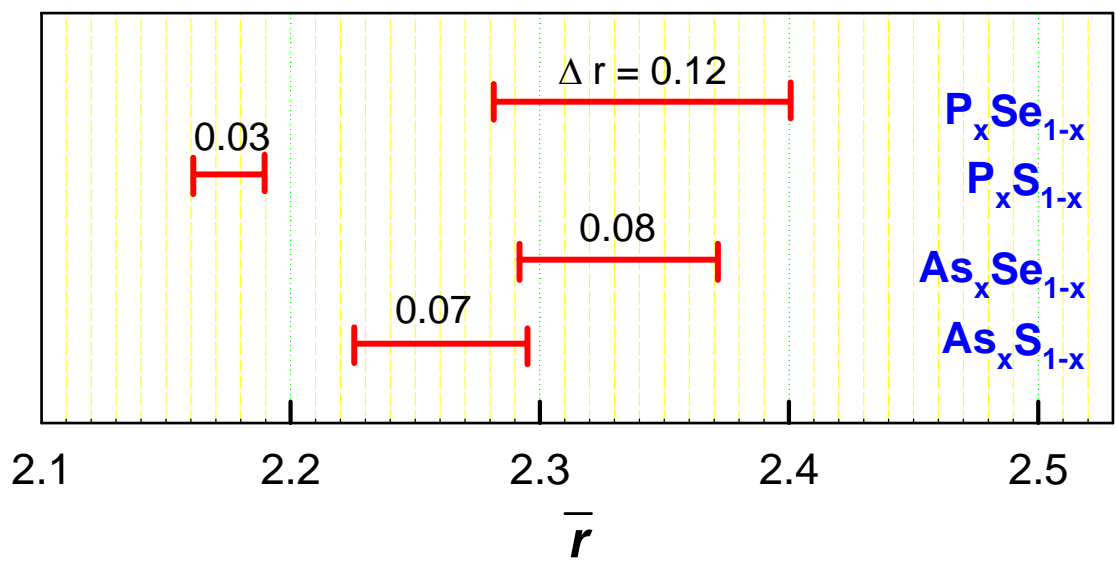

**Figure 19**